\documentclass[preprint,review,12pt]{elsarticle}

\usepackage{graphicx}
\usepackage{amssymb}
\usepackage{amsthm}
\usepackage{lineno}
\usepackage{amsmath}
\usepackage[super]{nth}

\journal{ASR}

\begin{document}

\begin{frontmatter}
 

\title{Comparative studies of Ionospheric models with GNSS and NavIC over the Indian Longitudinal sector during geomagnetic activities}



\author[a]{Sumanjit Chakraborty\corref{c-d54cc1eb1ca4}}
\ead{sumanjit11@gmail.com}\cortext[c-d54cc1eb1ca4]{Corresponding author.}
\author[a,b]{Abhirup Datta}
\ead{abhirup.datta@iiti.ac.in} 
\author[c]{Sarbani Ray}
\ead{sarbanir@yahoo.com }
\author[a]{Deepthi Ayyagari}
\ead{nagavijayadeepthi@gmail.com}
\author[c]{Ashik Paul} 
\ead{ap.rpe@caluniv.ac.in }

\address[a]{Discipline of Astronomy, Astrophysics and Space Engineering\unskip, IIT Indore\unskip, Simrol \unskip, Indore 453552, Madhya Pradesh, India}

\address[b]{Center for Astrophysics and Space Astronomy \unskip, Department of Astrophysical and Planetary Science \unskip, University of Colorado \unskip, Boulder \unskip, CO 80309, USA} 

\address[c]{Institute of Radio Physics and Electronics\unskip,University of Calcutta\unskip, Kolkata 700 009\unskip, West Bengal, India
\\\textbf{Accepted for publication in Advances in Space Research}
}

\begin{abstract} 

This paper presents the storm time comparative analysis of the performances of latest versions of global ionospheric models: International Reference Ionosphere (IRI) 2016, NeQuick 2 (NeQ) and the IRI extended to Plasmasphere (IRI-P) 2017 with respect to Navigation with Indian Constellation (NavIC) and Global Navigation Satellite System (GNSS) derived ionospheric Total Electron Content (TEC). The analysis is carried out under varying geomagnetic storm conditions during September 2017-November 2018, falling in the declining phase of solar cycle 24. TEC data from Indore, located near the northern crest of the Equatorial Ionization Anomaly (EIA) along with data obtained from the International GNSS Service (IGS) stations at Lucknow, located beyond the anomaly crest; Hyderabad, located between anomaly crest and magnetic equator and Bangalore, located near the magnetic equator have been analysed. 
The models generally overestimated during the storm periods with the exception of IRI-P, which matched (with an offset of about 3-5 TECU) with the enhancement observed on September 7, 2017 (during the strong storm of September 2017), from stations around the anomaly crest. No significant match was observed by the other two models. This match of IRI-P is attributed to the plasmaspheric contribution as well as the capability of assimilating measured TEC values into this model.
In the present study, to the best of our knowledge, first comparisons of the empirical model derived TEC with NavIC and GNSS measurements from an anomaly crest location, combined with the IGS observations from the magnetic equator to locations beyond the anomaly crest, are conducted during geomagnetically disturbed conditions.   
Since NavIC satellites are at higher altitudes($\sim$ 36000 km), the inclusion of NavIC data to the existing model could give better ionospheric predictions over the Indian subcontinent.

\end{abstract}


\end{frontmatter}


\newpage\section{Introduction}

Ionosphere, the ionized region of the atmosphere, extends from 60 km to beyond 1000 km above the surface of the Earth, with the equatorial region, confined within $\pm30^\circ$ magnetic dip, accounting for nearly one-third of the total global ionization. There is renewed interest in the upper atmospheric ionized regions primarily because of its deleterious effects on HF, VHF, and UHF radio communications and navigation and, aerospace, and the Global Navigation Satellite System (GNSS). As long-range radio waves are significantly affected by the ionosphere, constant observation of the ionospheric uncertainties is crucial for communication and navigational systems to work with higher efficiency \citep{sc:20}. A fundamental parameter to study the ionosphere is the ionospheric Total Electron Content (TEC) defined by the number of electrons integrated between two points, along a tube of unit cross-sectional area, expressed in TECU, where 1 TECU = 10$^{16}$ electrons/m$^2$. With the abundance of ionospheric data from the multitude of GNSS satellites, presently numbering more than 80 and projected to be in excess of 120 in the near future, coupled with the need for accurate navigation, various ionospheric models have been established which give predictions of the ionospheric TEC where actual data is absent. 

The equatorial and low-latitude ionosphere consists of several unique features such as the Equatorial Ionization Anomaly (EIA), the equatorial electrojet and the equatorial plasma irregularities, as a result of the horizontal orientation of the magnetic field at the geomagnetic equator, making this region to be the most dynamic and geosensitive \citep{sc:25,sc:22,sc:27,sc:21,sc:26,sc:23,sc:17,sc:2}. The electron density is expected to be maximum around the equator and decrease towards the poles as a result of sun's incident solar radiation over the equator (\citep{sc:1,sc:17,sc:2} and references therein). However, observations show this density to have peculiar crests around $\pm$15$^{\circ}$ magnetic latitudes and trough around the magnetic equator \citep{sc:1} and at the time of peak daytime electron density, a crest-to-trough density ratio of $\sim$1.6 that varies with varying geophysical conditions. This peculiarity is known as the EIA wherein plasma gets transported from the magnetic equator to the higher latitudes resulting in temporal and spatial variation in the ionospheric electron density and degrades TEC estimation capabilities of the empirical ionospheric models \citep{sc:36,sc:37,sc:16}. The ionosphere over central India falls within this anomaly crest region, where sharp latitudinal gradients of ionization are observed. 

The ionospheric TEC varies significantly during geomagnetic storms. These storms introduce temporary disturbances on the magnetosphere of Earth. They are caused by solar wind shock waves, which interact with the geomagnetic field \citep{sc:5}. The solar sources of such storms are the Coronal Mass Ejection (CME) and the Co-rotating Interaction Region (CIR), accompanied by the High Speed Solar Wind Streams (HSSWS). Generally, CIRs tend to occur during the solar minimum phase while the CMEs occurrence peak around solar maximum \citep{sc:33}. Whenever there are periods of such magnetic disturbances, the horizontal component of the Earth's magnetic field (H) gets perturbed, and the recovery to its average value is gradual. At mid-latitudes and the equatorial region, the decrease in H can be represented by a uniform magnetic field parallel to the geomagnetic dipole axis directed southward. The magnitude of this disturbance field, which is axially symmetric, varies with the storm-time or the time measured from the onset of the storm. This onset can be identified by a sudden increase in the value of H globally, known as the Storm Sudden Commencement (SSC). Following SSC, H remains above its average level for a few hours, known as the initial phase of the storm. It is followed by a substantial decrease in H, which indicates the main phase of the storm. The magnitude of this decrease in H indicates how severe the storm is. The disturbance field, represented by the Disturbance storm time (Dst) index, is axisymmetric with respect to the dipole axis (http://wdc.kugi.kyoto-u.ac.jp/dstdir/dst2/onDstindex.html). The severity of geomagnetic storms can be classified \citep{sc:8} as follows: 
\begin{itemize}
\item -30 nT $\leq$ Dst $<$ -50 nT signifies a weak storm
\item -50 nT $\leq$ Dst $<$ -100 nT signifies moderate storm
\item -100 nT $\leq$ Dst $<$ -200 nT signifies strong storm
\item -200 nT $\leq$ Dst $<$ -350 nT signifies severe storm
\item  Dst $\leq$ -350 nT signifies great storm
\end{itemize}
CME induced storms are generally strong and affect the Space Weather; however, even weak-to-moderate CIR induced storms have impacts on the ionosphere in a way similar to what a strong CME induced storm might have \citep{sc:40}.
Recently, the effects of the CME and CIR related strong geomagnetic storms on the ionization over Indian low-latitudes, in terms of the neutral dynamics, has been studied by \citep{sc:35}, thus highlighting the importance of studying the ionosphere during geomagnetically disturbed conditions, under low solar activity over the dynamic Indian longitude sector.

Studies on the performance of ionospheric models compared to real measured data observations from various satellite navigation systems have been performed by several researchers.  \citep{sc:18} assessed the performance of the older versions of the International Reference Ionosphere (IRI) 2012 and IRI-extended to Plasmasphere (IRI-P) 2015 over the equatorial and low-latitude regions of Africa and showed that IRI-P performed better compared to IRI model that underestimated the observed TEC. They further concluded that height limitation and inaccurate predictions of the electron densities of the IRI model created discrepancies in the observed and model data. Studies on the prediction capabilities of the NeQuick2 (NeQ) web model and IRI-P over the South American sector were performed by \citep{sc:34}, where they infer that model mismatch could be due to erroneous prediction of plasmaspheric contribution of TEC. \citep{sc:20} have compared the Ionolab derived GPS TEC over Istanbul, falling in the mid-latitude region, with the IRI 2016 and the IRI-P models, and concluded that IRI-P derived TECs are closer to the observed TEC compared to IRI. Comparison of International GNSS Service (IGS) VTEC with the IRI 2016 was made by \citep{sc:19} where they concluded that IRI derived TEC is consistent with the general trend of the ionosphere during low solar activity but vastly underestimated in low latitudes near EIA under high solar activity. \citep{sc:39} analyzed the performance of IRI 2016, IGS-GIM, and IRI-P 2017 with IGS GPS based observations during high solar activity period 2012-2015 and concluded that IRI-P presents better results compared to the other models but requires reliable performance during the disturbed periods. \citep{sc:38} evaluated the performance of IRI 2016 and IRI-P 2017 over central Asian mid-latitude regions and inferred that the models performed poorly during days of high solar irradiance. Most recently, \citep{sc:41} have investigated the possibilities of TEC usage over the low-latitude region and showed that the closest match is presented by IRI-P derived values.

Given the availability of a regional navigation satellite system like the Navigation with Indian Constellation (NavIC) along with the aid of the legacy GPS satellites, it becomes essential to look at various aspects of the ionosphere of the Indian longitude sector.
While several studies have shown the performance of ionospheric models with respect to real measured GPS data, geomagnetic storm time model deviations with respect to NavIC and GNSS (GPS, GLONASS and GALILEO) taken together over the Indian longitude sector, has not been reported extensively in the literature. In order to address this problem, the storm time performance of the latest ionospheric models: IRI 2016, NeQ and IRI-P 2017 have been evaluated with respect to NavIC and GNSS measured values. The period of analysis consists of strong, moderate, and weak storms spanning September 2017-November 2018, falling in the declining phase of the 2\nth{4} solar cycle. The stations considered for analysis are: Lucknow (26.91$^\circ$N, 80.95$^\circ$E geographic; magnetic dip 39.75$^\circ$N, located beyond the anomaly crest), Indore (22.52$^{\circ}$N and 75.92$^{\circ}$E geographic; magnetic dip: 32.23$^{\circ}$N, located near the anomaly crest), Hyderabad (17.41$^\circ$N, 78.55$^\circ$E geographic; magnetic dip 21.69$^\circ$N, located in between crest and magnetic equator) and Bangalore (13.02$^\circ$N, 77.5$^\circ$E geographic; magnetic dip 11.78$^\circ$N, located near the magnetic equator). 
The paper presents, for the first time, to the best of our knowledge, evaluation of ionospheric model derived TEC, covering a large spatial distribution over the Indian longitude sector, under variable geomagnetic conditions, during the declining phase of solar cycle 24.

\section{The Ionospheric models}

The IRI 2016 is an empirical model of the ionosphere. The sources of data to this model are the incoherent scatter radars and the dense worldwide network of ionosondes along with the Alouette topside sounder's in-situ instruments on board satellites. 
The inputs to this model are the day of the year, the geographic/geomagnetic latitude, longitude, and altitude. To name a few, the model output provides the electron temperature and density, ion temperature, and composition and the TEC from 60-2000 km altitude range \citep{sc:14}.
The NeQ is an upgraded version of the NeQuick model. This model  uses a modified Di Giovanni and Radicella (DGR) profile formulation \citep{sc:4} that consists of five semi-Epstein layers \citep{sc:13} with modelled thickness parameters \citep{sc:12}, for describing the ionospheric electron density from 90 km to the maximum height of the F2 layer. The model topside is represented by a semi-Epstein layer with a height-dependent thickness parameter that is determined empirically \citep{sc:6},\citep{sc:3}. The inputs to this model are the time, the position (geographic latitudes and longitudes), and either the F10.7 solar radio flux or the daily sunspot number while the outputs are the electron concentration corresponding to the user input location and time. Specific routines are present in NeQ to evaluate electron density and the corresponding TEC by the method of numerical integration \citep{sc:11}.
The IRI-P 2017 \citep{sc:15} is the IRI extended to the plasmasphere, which consists of the most developed plasmaspheric and ionospheric model. The model presents a better representation of the ionosphere due to the input of real values into the model \citep{sc:28}. The model additionally has a scale height parameter which determines the structure of the IRI topside electron density profile \citep{sc:29}. 
The inputs of this model are the day of year, latitude, longitude with the F10.7 solar radio flux, or the daily sunspot number. The outputs of the model are the maximum height of the F2 layer (hmF2), critical frequency of F2 layer (foF2), ionospheric TEC and, electron density, among others.

\section{The NavIC}

NavIC, previously known as the Indian Regional Navigation Satellite System (IRNSS), is a regional satellite navigation system developed by ISRO.
The space segment consists of three Geostationary Earth Orbit (GEO) and three Geosynchronous Orbit (GSO) satellites. The position of these satellites are such that all of them have continuous radio visibility with the Indian control stations. These satellites broadcast signals at 24 MHz bandwidth of spectrum in the L5 and S band having carrier frequencies 1176.45 MHz and 2492.03 MHz, respectively \citep{sc:10}. 

\section{Data and Methodology} 

\begin{figure*}
\includegraphics[width=\columnwidth,height=4.5in]{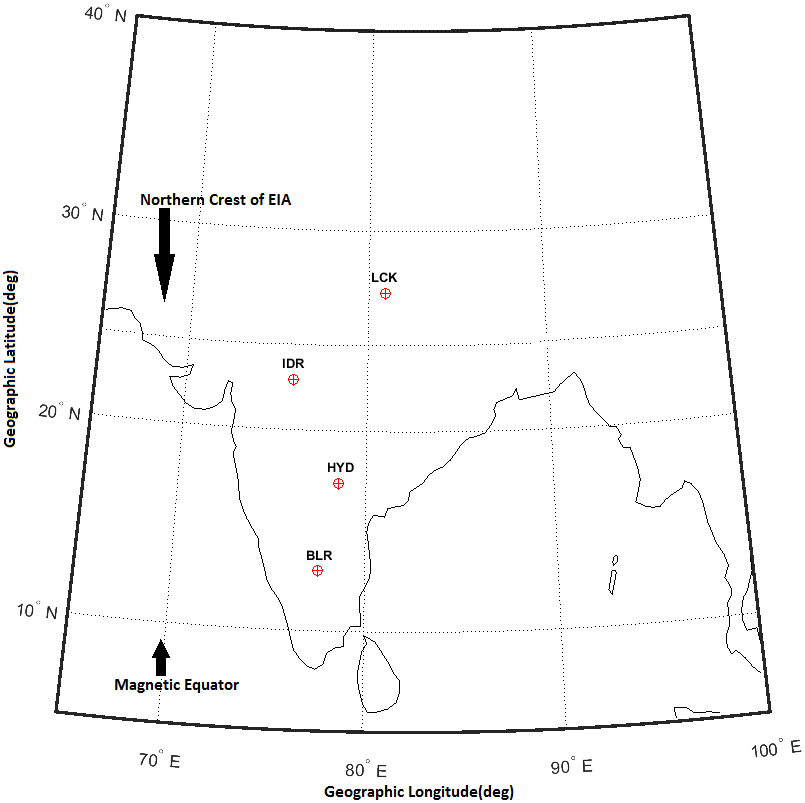}
\caption{The map of India depicting the geographic locations of the NavIC and GNSS receivers over Indore(IDR) and IGS-GPS receivers over Lucknow(LCK), Hyderabad(HYD) and Bangalore(BLR). The locations of the magnetic equator and northern crest of EIA are also indicated in the map.}
\label{sc1}
\end{figure*}

A multi-constellation(GPS, GLONASS and GALILEO) and multi-frequency (GPS L1, L2 and L5) GNSS receiver along with a NavIC receiver (provided by the Space Applications Centre (SAC), ISRO) capable of receiving L5 and S1 signals along with GPS L1 signal, are operational in the Discipline of Astronomy, Astrophysics and Space Engineering of Indian Institute of Technology, Indore (IITI). In order to avoid the sharp latitudinal spatio-temporal gradient in the electron density that exists in and around the anomaly region, the Slant TEC (STEC) obtained at the output of the receivers are converted to the equivalent Vertical TEC (VTEC) using: 
\begin{equation}
VTEC = \frac{STEC}{M_F}
\end{equation}
where $M_F$ is the mapping or obliquity factor (\citep{sc:30,sc:42,sc:7} and references therein) given by:
\begin{equation}
M_F = \Bigg[1 - \Big(\frac{{R_e} cos{E}}{R_e+h}\Big)^{2}\Bigg]^{-1/2}
\end{equation}
where $R_e$ is the radius (6371 km) of the Earth, $h$ the altitude of the ionosphere that is considered as a thin shell at 350 km and $E$ the satellite's elevation angle.
VTEC data have also been analysed and obtained from the IGS stations at Lucknow (26.91$^\circ$N, 80.95$^\circ$E geographic; magnetic dip 39.75$^\circ$N), Hyderabad (17.41$^\circ$N, 78.55$^\circ$E geographic; magnetic dip 21.69$^\circ$N) and Bangalore (13.02$^\circ$N, 77.5$^\circ$E geographic; magnetic dip 11.78$^\circ$N), available at the Scripps Orbit and Permanent Array Center (SOPAC) website (http://sopac-csrc.ucsd.edu/index
.php/sopac). The elevation cut-off chosen for analyzing the VTEC, for the GNSS constellation, is 20$^\circ$ to avoid the effect of multipath on the receiving signals. The geographic location of Indore and the three IGS stations (Lucknow, Hyderabad and Bangalore) spanning over the Indian subcontinent is shown in Figure \ref{sc1}.

The hourly Dst (nT) and the K$_p$ indices are obtained from the World Data Center for Geomagnetism, Kyoto (http://wdc.kugi.kyoto.ac.jp). The daily sunspot number (SSN) is obtained from (http://www.sidc.be/
silso/datafiles). The solar radio flux (F10.7, s.f.u), the Auroral Electrojet (AE, nT), the Interplanetary Magnetic Field (IMF,Bz, nT) and Interplanetary Electric Field (IEF,Ey, mV/m) are obtained from NASA's Space Physics Data Facility (SPDF) omniweb service (http://omniweb.gsfc.nasa.gov) at 1 minute resolution and have been analysed for the storm periods. Additionally, the ionospheric model derived TEC data during the storm periods are obtained from IRI (https://ccmc.gsfc.nasa.gov/modelweb/models/
iri2016\_vitmo.php), NeQ (https://t-ict4d.ictp.it/nequick2/nequick-2-web-model) and from the IRI-P (http://www.ionolab.org) models.
Table \ref{sct} shows the minimum Dst values along with the corresponding day and time for the selected storms in addition to the SSN, F10.7, K$_p$ and AE indices. 
\begin{table*}
\centering
\begin{center}
\caption{Minimum Dst values with the corresponding time and type of storms in addition to the daily SSN, F10.7 solar radio flux, the K$_p$ and AE indices. The dashed lines indicate unavailable data during that period.}
\resizebox{\textwidth}{0.25\textheight}{
\begin{tabular}{|l|c|c|l|c|c|c|c|c|c|c|}
\hline 
Date & SSN & F10.7(s.f.u) & K$_p$ & Dst(nT) & AE(nT) & UT(hh:mm) & Type & Source \\ 
\hline \hline
Sep 08, 2017 & 88 & 118.5 & 7- & -124 & 791 & 02:00 & strong & CME  \\ 
Sep 28, 2017 & 42 &  91.2 & 5 & -55  & 789 & 07:00 & moderate & CIR  \\
Oct 14, 2017 & 11 &  68.6 & 5 & -57  & 741 & 06:00 & moderate & CIR   \\
Dec 04, 2017 & 00 &  66.4 & 4 & -45  & 292 & 22:00 & weak & CIR     \\
Jan 19, 2018 & 12 &  68.5 & 4 & -30  & 220 & 09:00 & weak & CIR    \\
Feb 23, 2018 & 00 &  68.2 & 4 & -31  & 500 & 12:00 & weak & CIR   \\
May 06, 2018 & 15 &  68.4 & 5 & -56  & --- & 02:00 & moderate & CIR \\
Oct 07, 2018 & 00 &  69.4 & 5 & -53  & --- & 22:00 & moderate & CIR \\
Nov 05, 2018 & 00 &  67.2 & 5 & -53  & --- & 06:00 & moderate & CIR \\
\hline 
\end{tabular}}
\end{center}
\label{sct}
\end{table*}

\newpage
\section{Results and Discussions}

In this section, the performance of the three models (IRI-P, NeQ and IRI) have been evaluated with respect to real measured data (NavIC and GNSS) observed from the stations: Lucknow, Indore, Hyderabad and Bangalore that ranges from beyond the EIA to near the magnetic equator over the Indian longitude sector. The strong, moderate and weak storms during the period of September 2017-November 2018, falling in the declining phase of solar cycle 24, have been analyzed. As sample cases, storms of September 8, 2017, January 19, 2018 and November 5, 2018 are discussed in the following subsections. 

\subsection{Strong storm of September 8, 2017}

\begin{figure*}
\includegraphics[width=\columnwidth,height=4.5in]{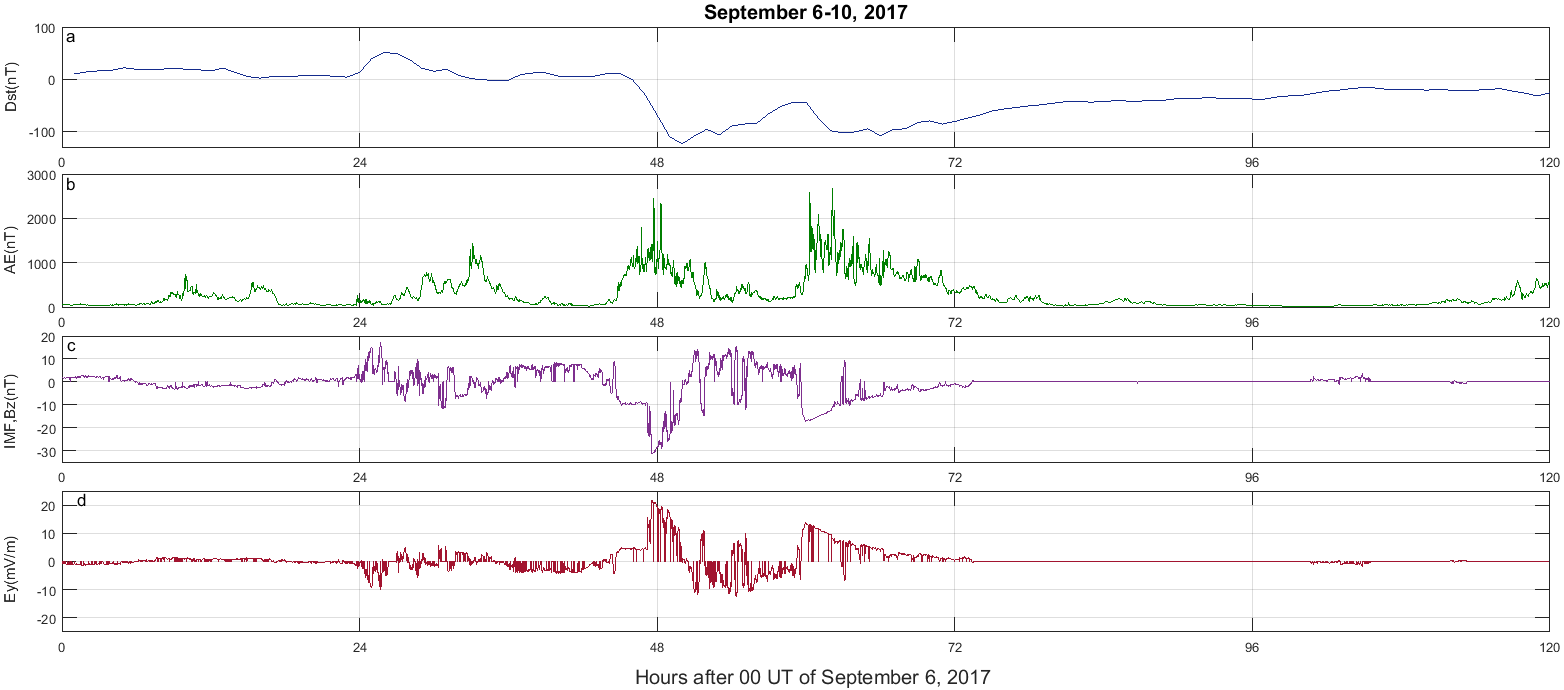}
\caption{Dst, AE and interplanetary parameters for September 6-10, 2017. September 8 (48-72 UT(h) in the plot) signifies the day of Dst minimum.}
\label{sc2}
\end{figure*}

As a result of the arrival of a CME on September 6, 2017, a G4 level (K$_p$=8, severe) geomagnetic storm, according to the National Oceanic and Atmospheric Administration (NOAA) Space Weather Scales 
(https://www.
swpc.noaa.gov/), was observed at 23:50 UT on September 7, 2017 followed by two more at 01:51 UT and 13:04 UT on September 8, 2017.
Figure \ref{sc2} shows the variation of Dst index along with the interplanetary parameters for the period September 6-10, 2017. In Figure \ref{sc2}a, a SSC can be observed when there was a sudden increase in Dst with a value of 22 nT at 05:00 UT on September 7. Dst reached a minimum value of -124 nT at 02:00 UT and again -109 nT at 18:00 UT on September 8 signifying the strong storm of double-peak. Figure \ref{sc2}b shows the AE values that first peaked to 2447 nT at 23:43 UT on September 7 and had another peak with a higher magnitude and a value of 2677 nT at 14:07 UT on September 8, signifying heat generation due to highly perturbed electric fields at high latitudes. This higher magnitude of the second peak of AE is in accordance with the Dst which also showed a second dip at 18:00 UT. Figure \ref{sc2}c and \ref{sc2}d show the IMF, Bz and IEF, Ey, having variations in the opposite direction, with a minimum value of -31.21 nT and a maximum value of 21.68 mV/m respectively at 23:32 UT.  

\begin{figure*}
\includegraphics[width=\columnwidth,height=4.5in]{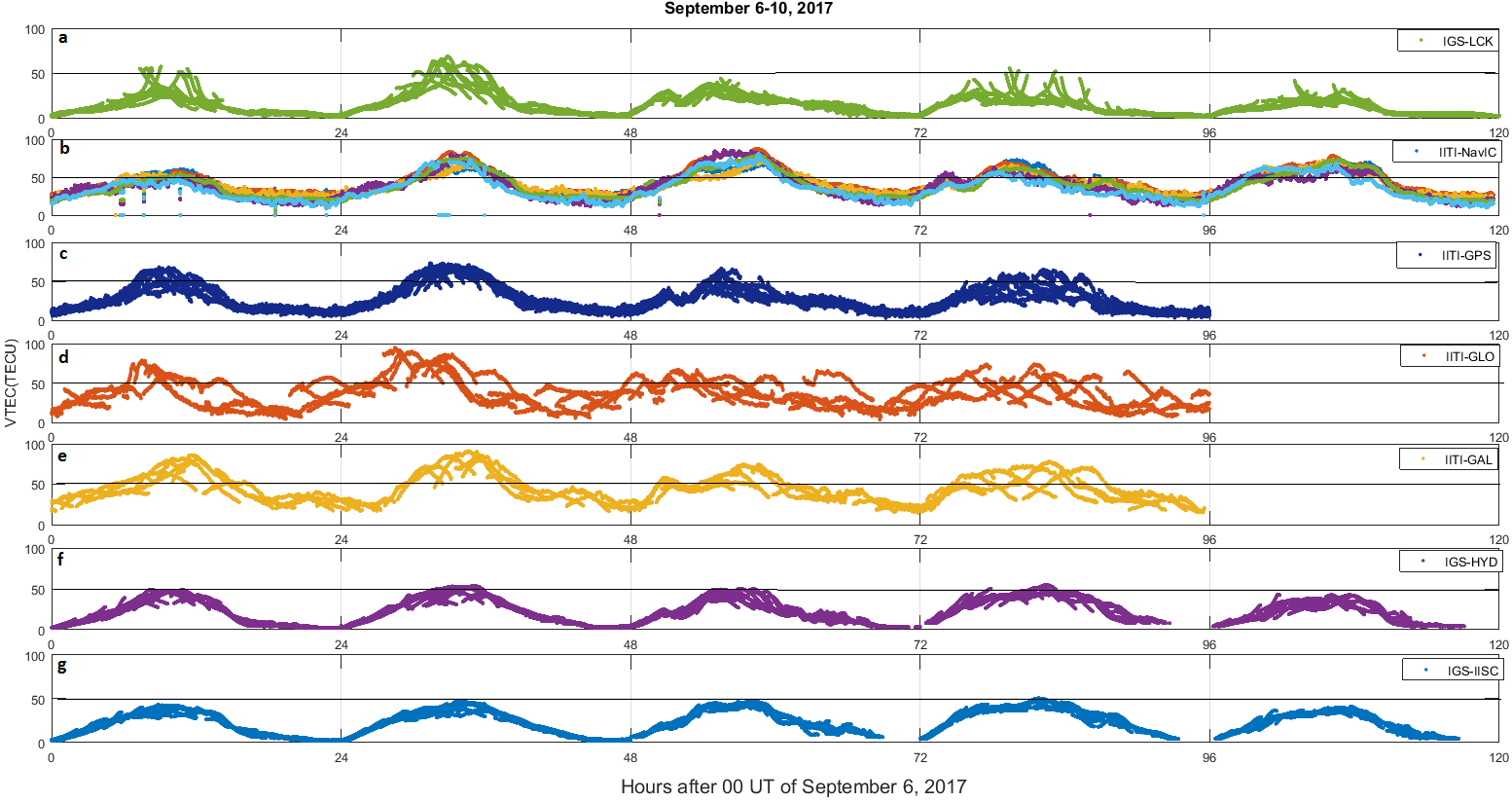}
\caption{Diurnal variation of VTEC during September 06-10, 2017 observed from a distributed chain of four stations in the Indian longitude sector.}
\label{sc3}
\end{figure*}

Diurnal variations of VTEC obtained from IGS GPS observables from the stations Lucknow, Hyderabad, Bangalore along with NavIC, GPS, GLONASS and GALILEO observables from Indore at a resolution of 1 minute is shown in Figure \ref{sc3}. Surprisingly, a higher value of VTEC is observed on September 7 while a lower value is observed on September 8, i.e the day when Dst dropped to a minimum, by IGS GPS at Lucknow and the GNSS observables from Indore. This signifies a decrease in TEC on September 8 over Lucknow and Indore
which are beyond and near the anomaly crest respectively, whereas no significant changes in TEC due to the storm is observed over the stations Hyderabad and Bangalore located closer to the magnetic equator. This shows that during storm time conditions, ionosphere in and around the anomaly crest become perturbed in comparison to that near the magnetic equator or the anomaly trough as because the equatorial fountain intensifies during geomagnetic storms resulting in strengthening the EIA crest. 

\begin{figure*}
\includegraphics[width=0.5\columnwidth,height=2.5in]{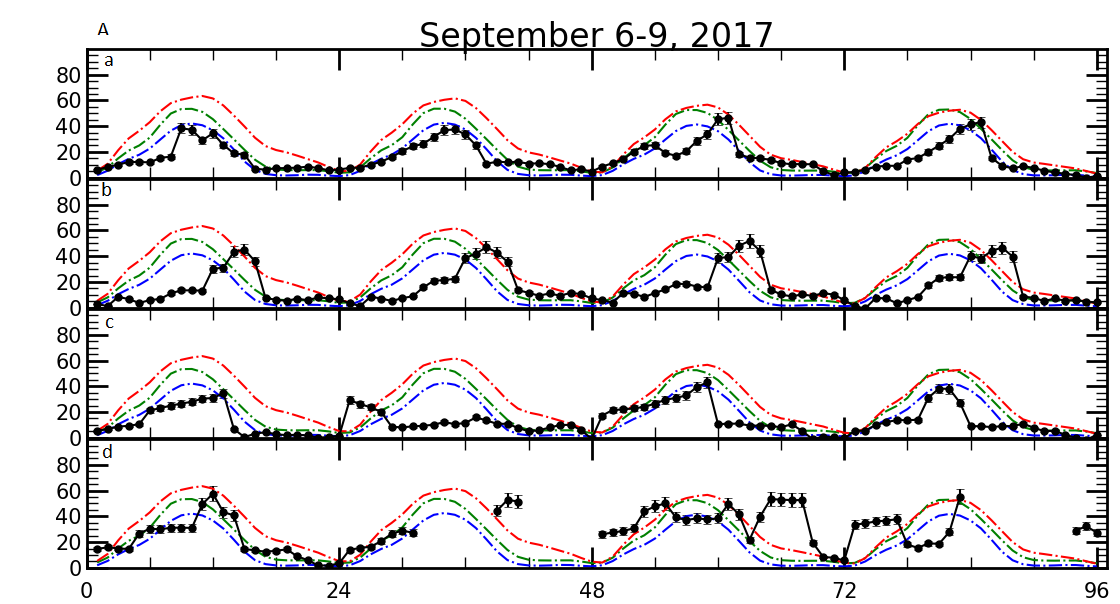}
\includegraphics[width=0.5\columnwidth,height=2.5in]{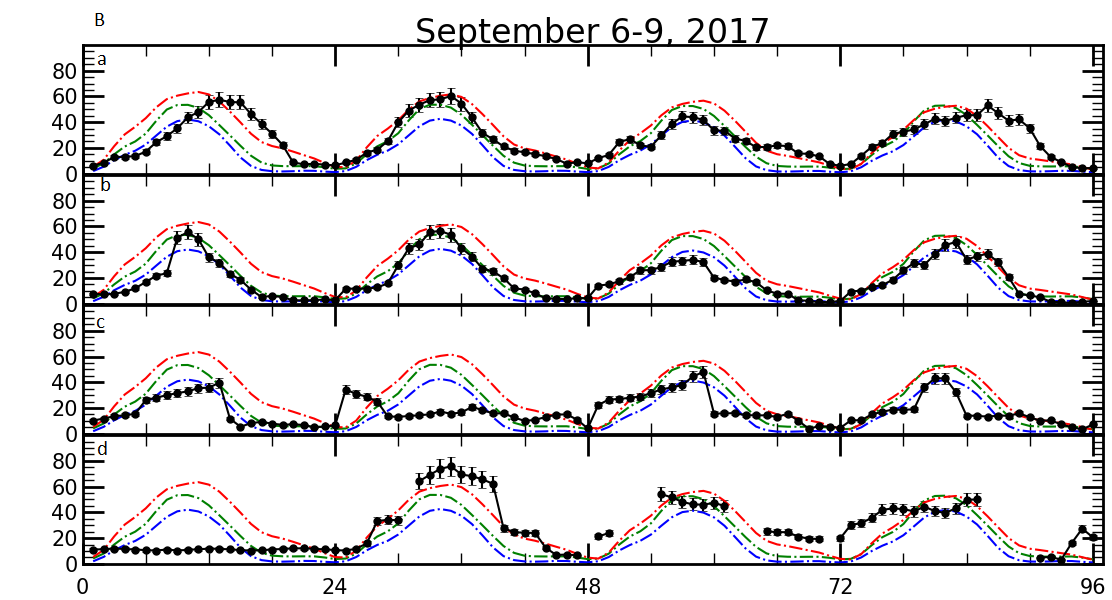}\\
\includegraphics[width=0.5\columnwidth,height=2.5in]{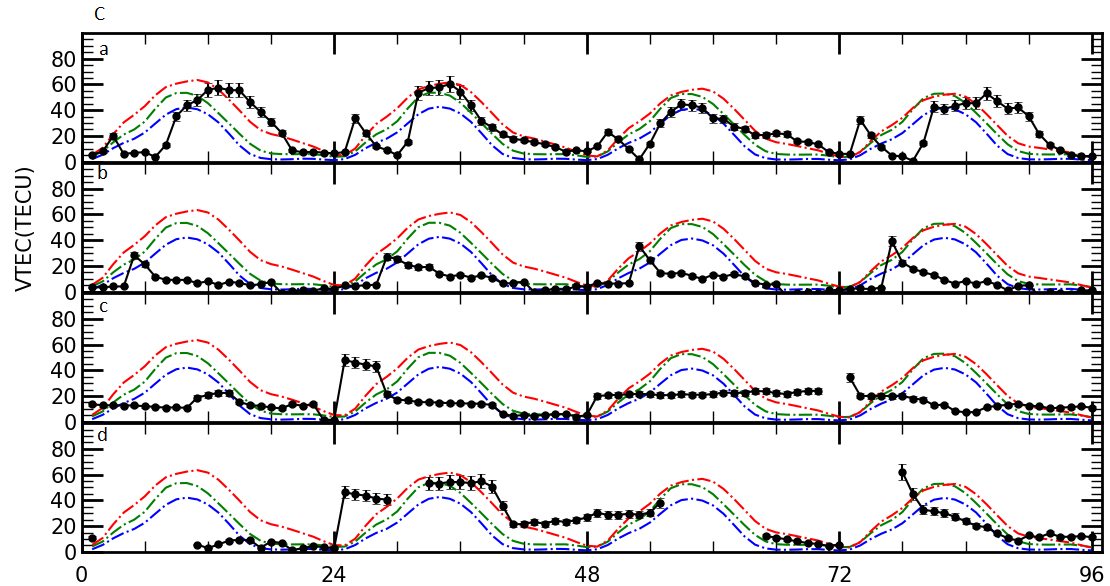}
\includegraphics[width=0.5\columnwidth,height=2.5in]{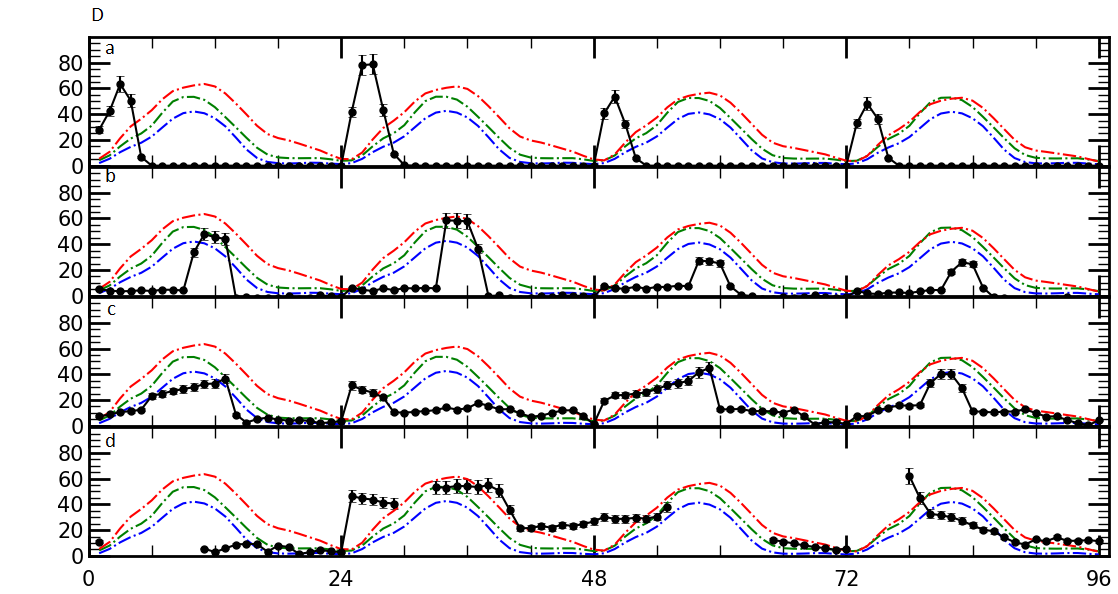}\\
\includegraphics[width=0.5\columnwidth,height=2.5in]{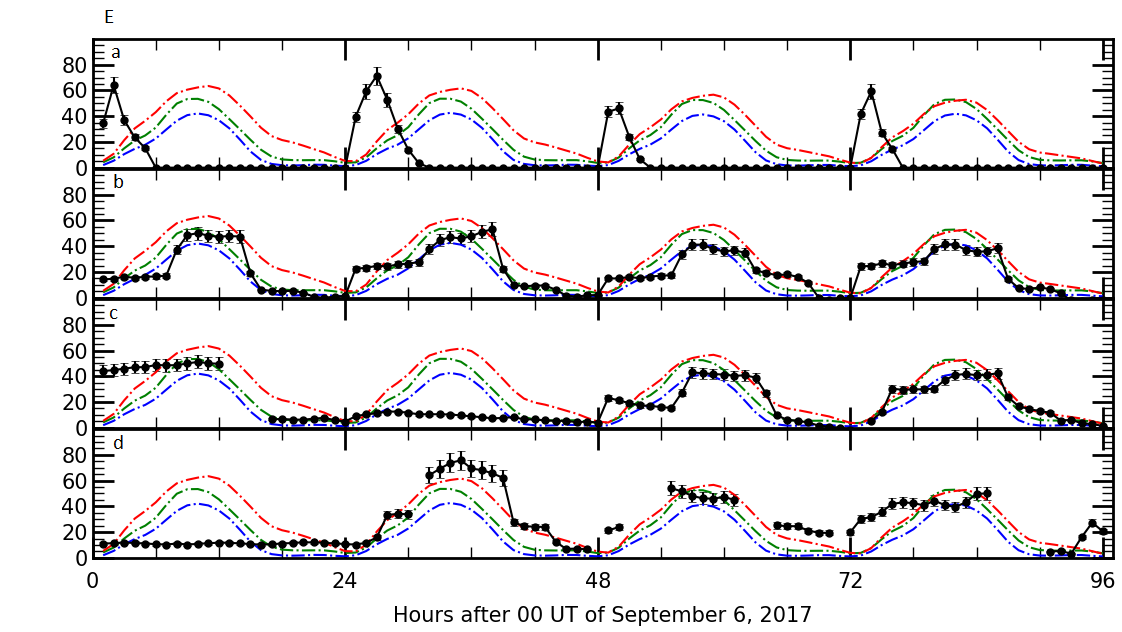}
\includegraphics[width=0.5\columnwidth,height=2.5in]{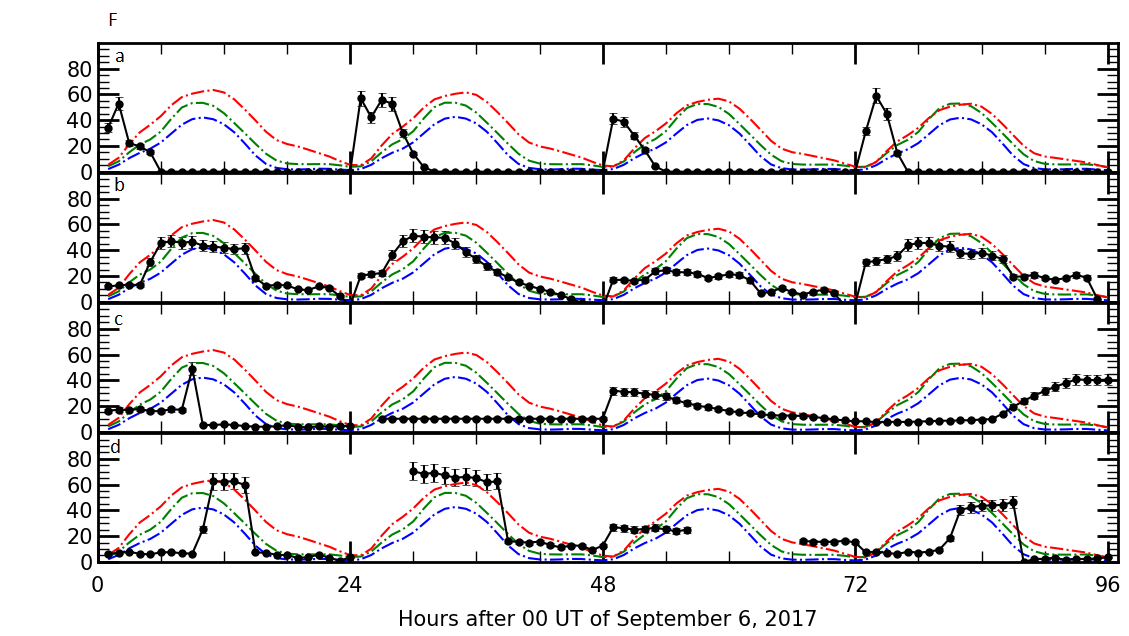}\\
\caption{Diurnal variations of model derived VTEC from IRI-P(green), NeQ(red) and IRI(blue) are compared with PRNs 2-7(panels A-F respectively) of NavIC(subplots:(a)) and all PRNs of GPS(subplots:(b)), GLONASS(subplots:(c)) and GALILEO(subplots:(d)), over Indore during September 5-9, 2017. One sigma error-bar of the measured values(black) are also shown for the period.}
\label{sc4}
\end{figure*}

Figure \ref{sc4} shows the comparative analyses of IRI, NeQ and IRI-P models with observations along with a one sigma error-bar from NavIC, GPS, GLONASS and GALILEO over Indore, depicting a multi constellation picture of the ionosphere during this period. In this figure, the available GEO and GSO satellites of NavIC, namely PRNs 2-7, are depicted in panels A-F respectively. The GPS, GLONASS and GALILEO satellites which are within the 2$^\circ\times 2^\circ$ IPP reception zone of the NavIC satellites, are taken for analysis. It can be observed from Figure \ref{sc4}A that the model derived TEC are overestimating the measured values as well as they are unable to capture the true variation during disturbed conditions. 
From Figure \ref{sc4}B a shift in the diurnal value is observed by NavIC (panel a) which is not captured by the models, however there is a close match ($\sim$ 4-5 TECU) on September 7, 2017 (one day prior to the storm day) by IRI-P (green line) with respect to NavIC and GPS (panel b) measurements.
Similar variation can be observed from Figure \ref{sc4}C which shows the model values being unable to replicate the measured variations.
Figures \ref{sc4}D-F clearly shows the inability of the models to capture real time variation of NavIC (panel a) while a close variation is observed with respect to GPS values (panel b).
From the overall observations, it can be inferred that IRI-P which accounts for the plasmaspheric contribution towards electron density presents the nearest match with NavIC and GPS measured TEC over a location near the anomaly crest during the disturbed period.   

\begin{figure*}
\includegraphics[width=\columnwidth,height=4.5in]{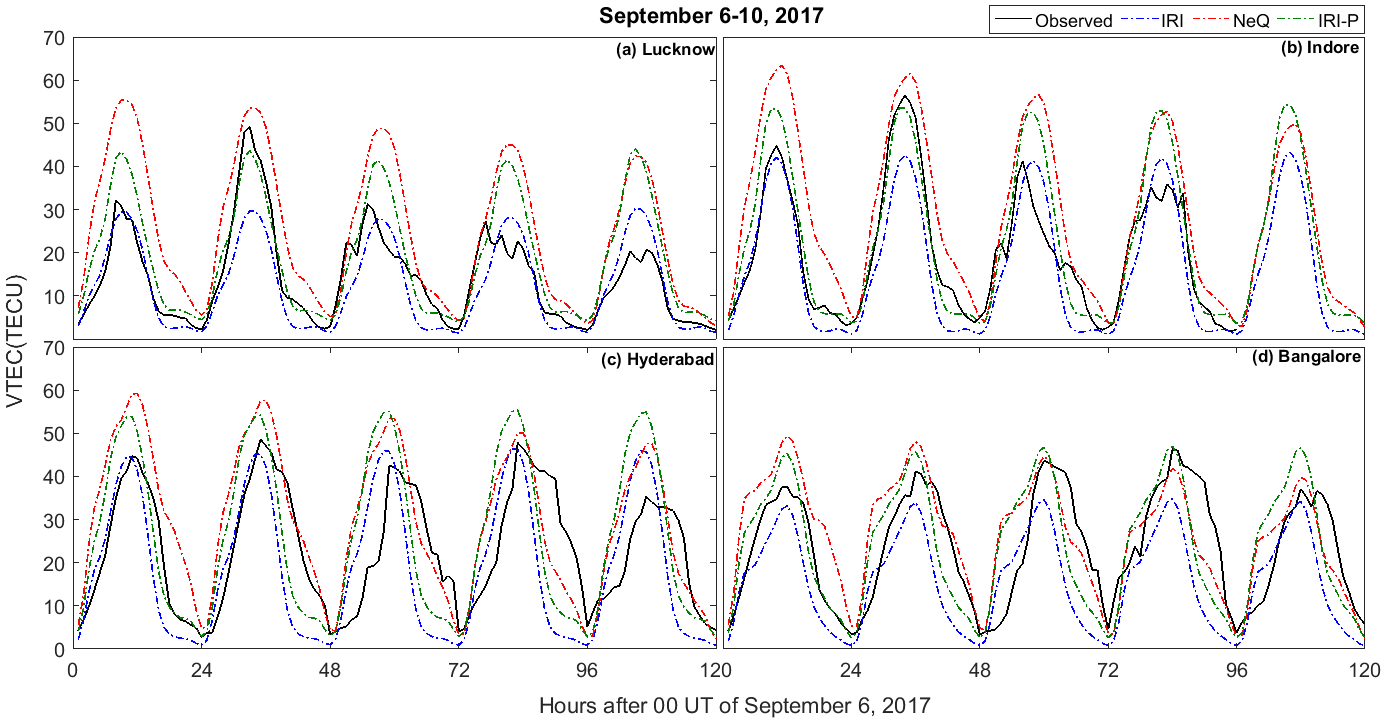}
\caption{Diurnal variations of VTEC from observed values of GPS are compared with model derived values of IRI(blue), NeQ(red) and IRI-P(green) over (a) Lucknow, (b) Indore, (c) Hyderabad and (d) Bangalore during the period of September 6-10, 2017.}
\label{sc5}
\end{figure*}

Figure \ref{sc5} shows the latitudinal variation of diurnal VTEC obtained from IRI-P, NeQ, IRI and GPS over the stations:(a) Lucknow,(b) Indore,(c) Hyderabad and (d) Bangalore during the period of September 6-10, 2017 thus covering a large spatial distribution ranging from beyond the northern anomaly crest to the magnetic equator over the Indian subcontinent, under strong storm time conditions. 
From Figure \ref{sc5}a, it is observed that over Lucknow, IRI TEC matches with GPS TEC on September 6 while the other two models overestimate. The enhancement in TEC observed on September 7 is captured by IRI-P with a difference, in the diurnal maximum between GPS and IRI-P values, of about 3-4 TECU, but gets overestimated by NeQ and largely underestimated by IRI, thus indicating to a better predictive capability of the IRI-P model. On the day of the storm and the day after, a close match in TEC is observed by IRI but underestimated by NeQ and IRI-P. 
Similar pattern is observed over Indore but with higher values, as shown in Figure \ref{sc5}b, by both the models and real data. This is expected as Indore is nearest to the anomaly crest. On September 6, IRI matches with real data. The enhancement on September 7 is again well matched (with an offset of $\sim$4-5 TECU at diurnal maximum) by IRI-P. 
Figure \ref{sc5}c shows a close match of IRI with GPS TEC on September 6 but is unable to capture the TEC decrease on September 8. However, in Figure \ref{sc5}d, IRI underestimates observed TEC on all days while NeQ and IRI-P overestimates on September 6 and 7.
In general, the IGS GPS plots clearly confirm gradual increase in the TEC magnitude as one goes northward from Bangalore (nearest to the magnetic equator) to Hyderabad (an intermediate location) to Indore (nearest to EIA crest). The values then decrease at Lucknow, located beyond the northern crest.

Thus in Figure \ref{sc4}, a multi constellation analysis has been carried out over a location near the EIA while Figure \ref{sc5} portrays a single constellation's observations over multiple locations in order to bring out the storm time variability of ionospheric TEC over a spatial distribution from beyond the EIA to near the magnetic equator of the Indian region.     
Furthermore, both the figures show enhancements in the observed TEC on September 7 over Lucknow and Indore located near the anomaly crest and are well captured by IRI-P while almost no variation is observed over Hyderabad and Bangalore which are not that well captured by the models.

\newpage\subsection{Weak storm of January 19, 2018}

\begin{figure*}  
\includegraphics[width=\columnwidth,height=4.5in]{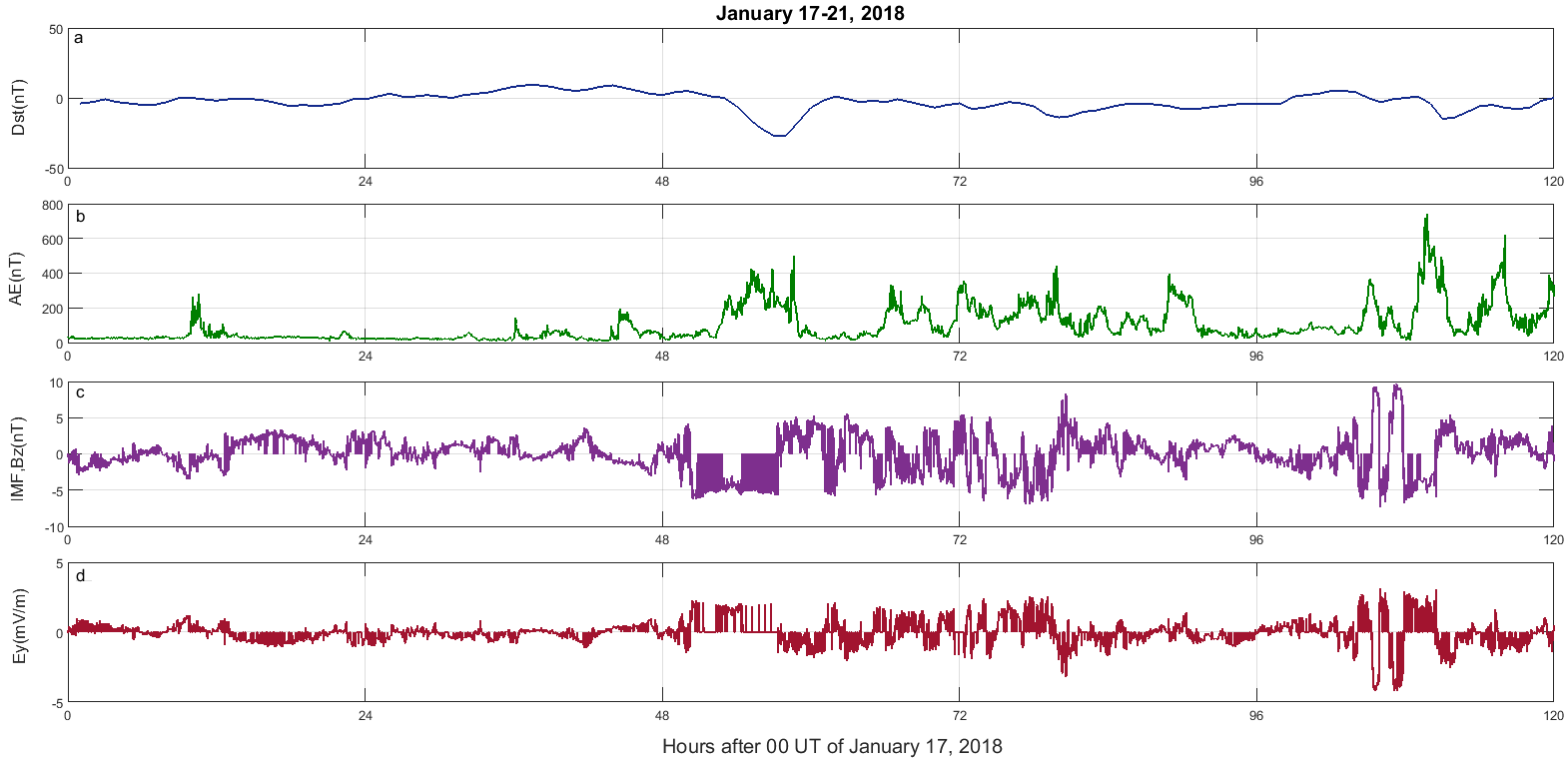}
\caption{Dst, AE and interplanetary parameters for January 17-21, 2018. January 19(48-72 UT(h) in the plot) signifies the day of Dst minimum.}
\label{sc6}
\end{figure*}

A CIR that originated from a positive polarity coronal hole, accompanied by HSSWS, hit the Earth's magnetic field on January 14, 2018. This event sparked a G1 level (K$_p$=5, minor) geomagnetic storm as reported by NOAA.
Figure \ref{sc6}, in a way similar to Figure \ref{sc2}, shows the Dst and interplanetary parameters' variation over the selected period. Figure \ref{sc6}a shows the Dst reaching a minimum with a value of -30 nT at 09:00 UT on January 19 thus signifying the storm to be weak in nature. Figure \ref{sc6}b shows the AE values which was 499 nT at 10:38 UT on January 19 around the time of Dst minimum. AE had a second peak with a higher value of 739 nT at 13:44 UT and a third peak having a value of 621 nT at 20:01 UT on January 21. Figures \ref{sc6}c and \ref{sc6}d show the IMF, Bz and IEF, Ey with a minimum value of -6.87 nT and a maximum value of 2.94 mV/m respectively at 11:54 UT on January 21. 

\begin{figure*}
\includegraphics[width=\columnwidth,height=4.5in]{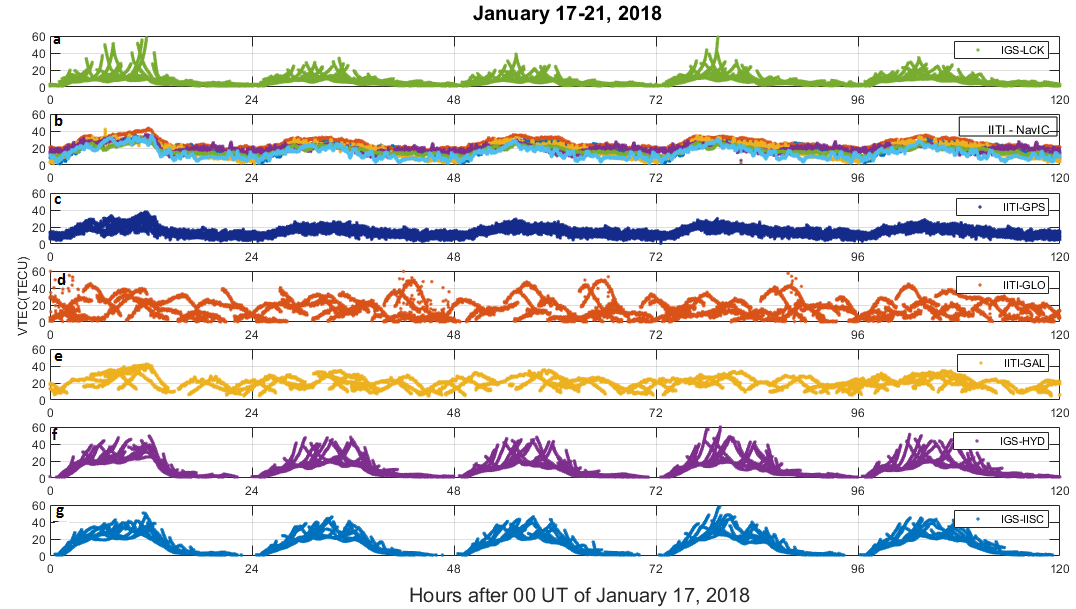}
\caption{Diurnal variation of VTEC during January 17-21, 2018 observed from a distributed chain of four stations in the Indian longitude sector.}
\label{sc7}
\end{figure*}

\begin{figure*}
\includegraphics[width=0.5\columnwidth,height=2.5in]{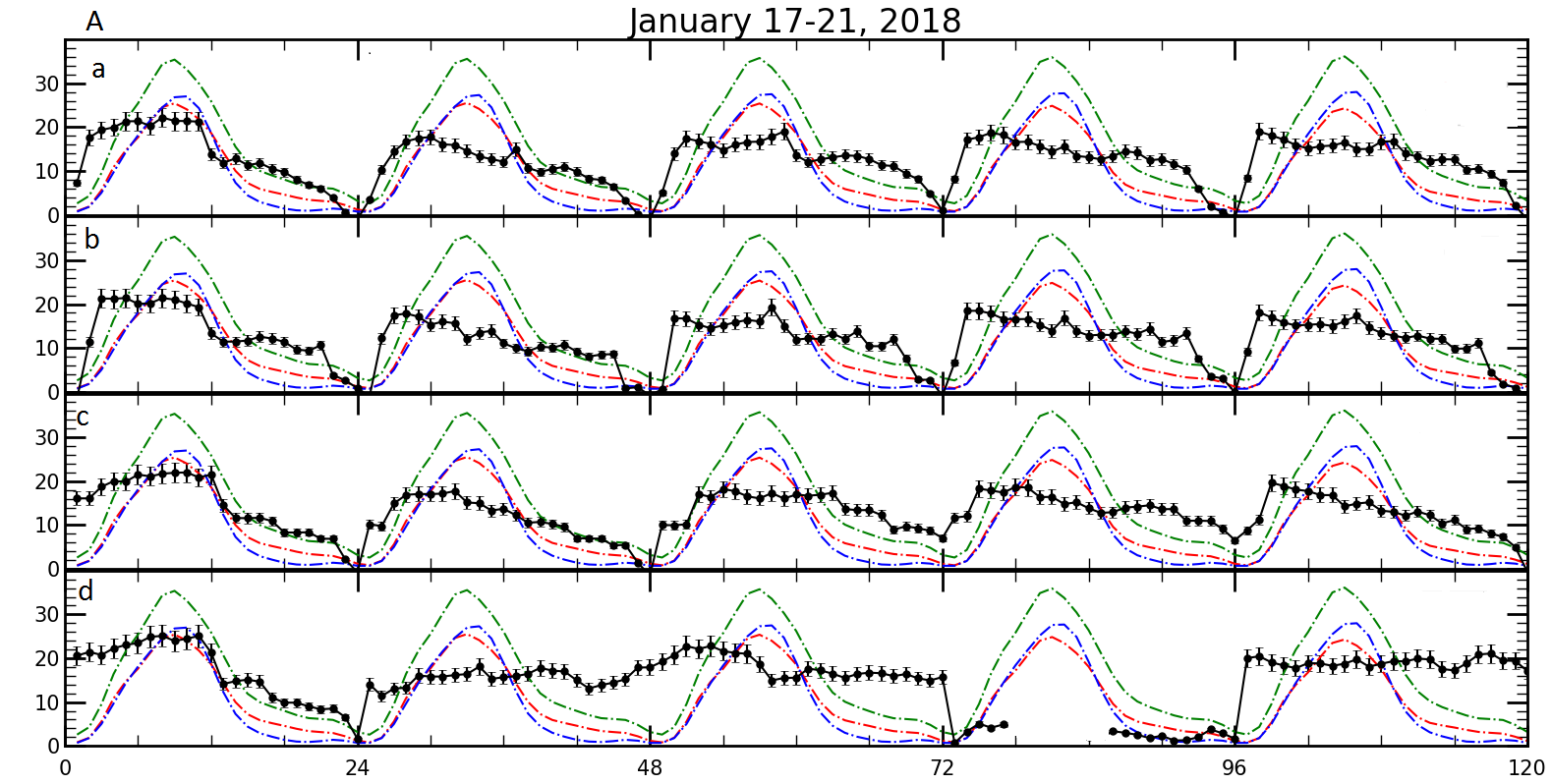}
\includegraphics[width=0.5\columnwidth,height=2.5in]{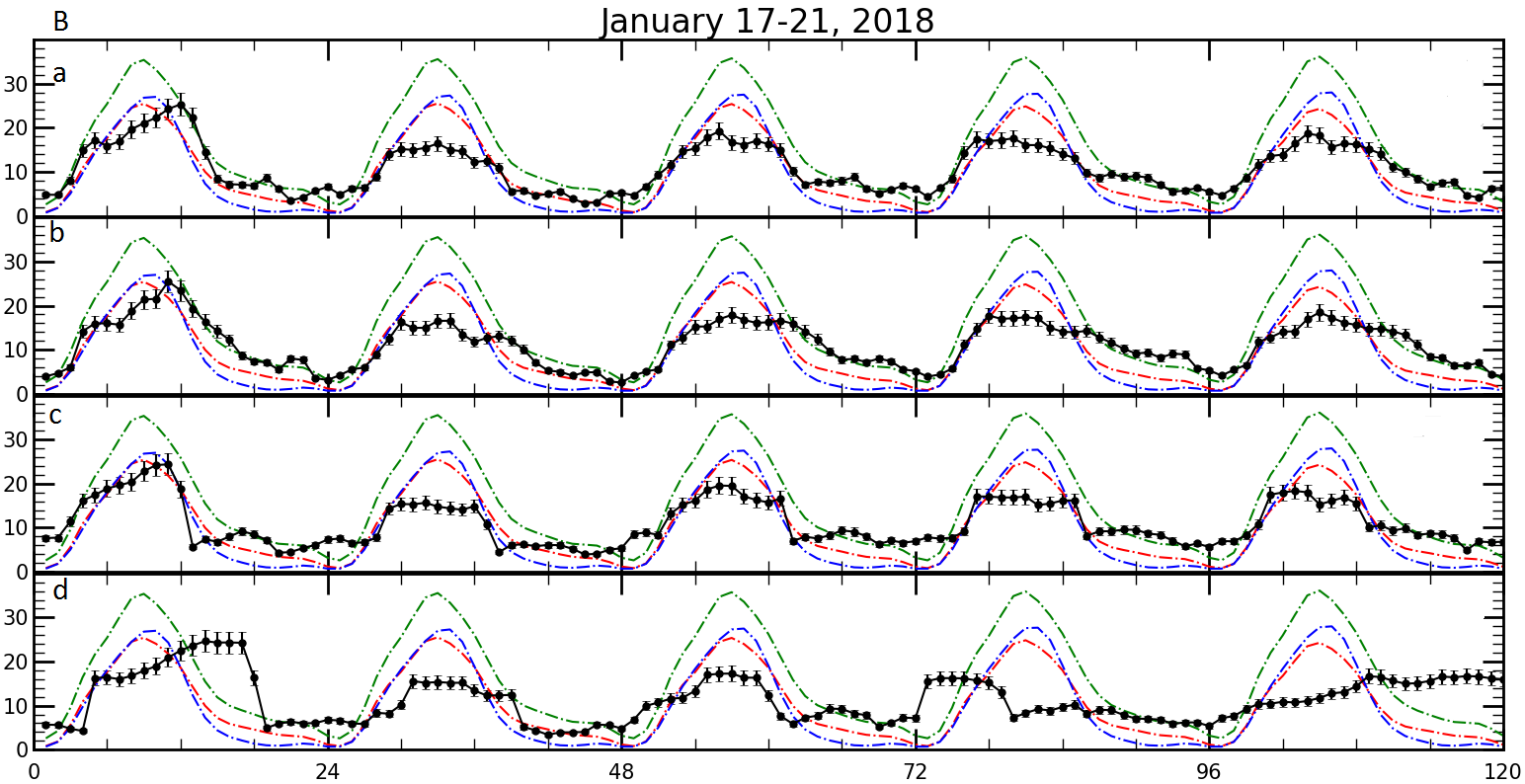}\\
\includegraphics[width=0.5\columnwidth,height=2.5in]{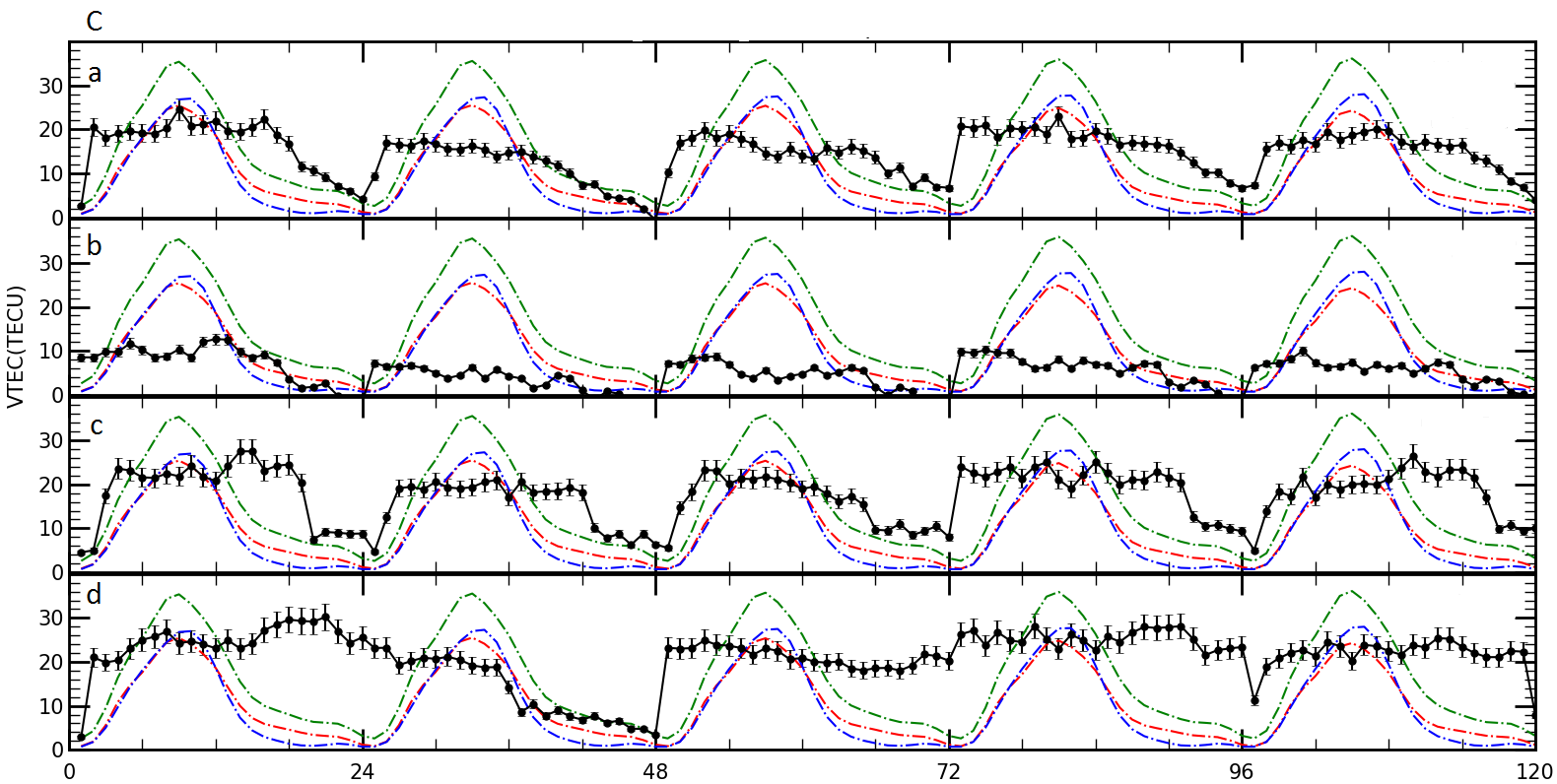}
\includegraphics[width=0.5\columnwidth,height=2.5in]{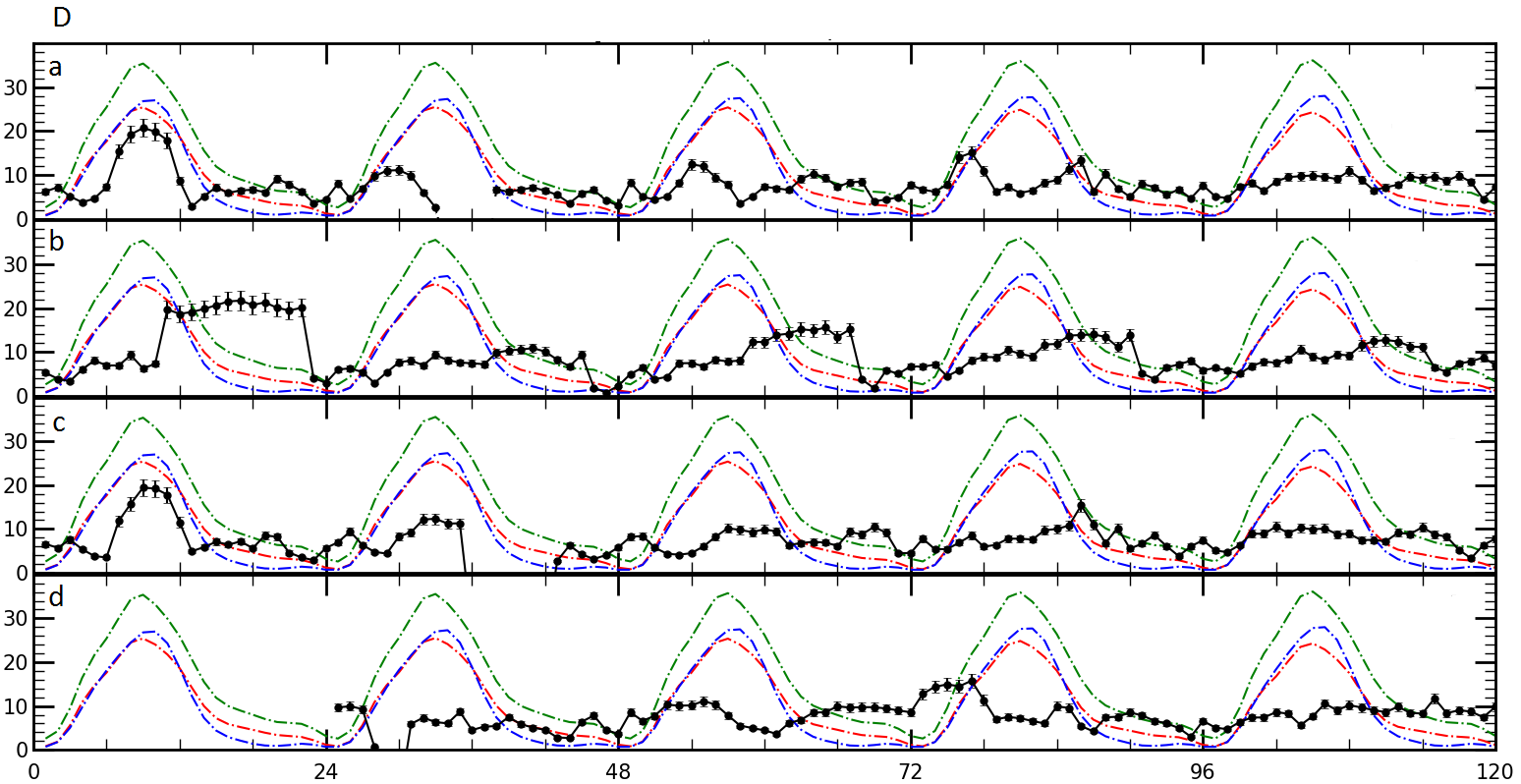}\\
\includegraphics[width=0.5\columnwidth,height=2.5in]{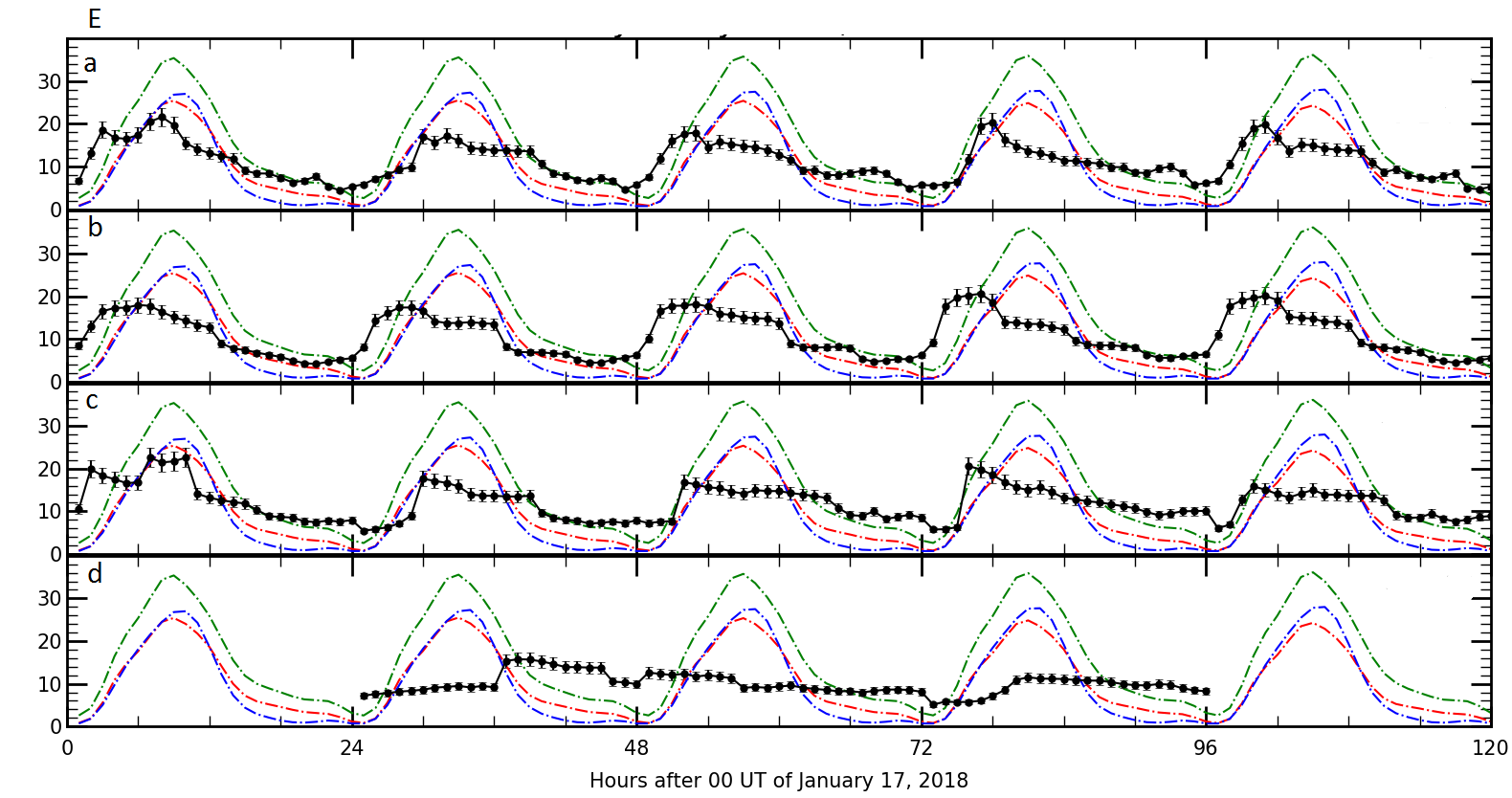}
\includegraphics[width=0.5\columnwidth,height=2.5in]{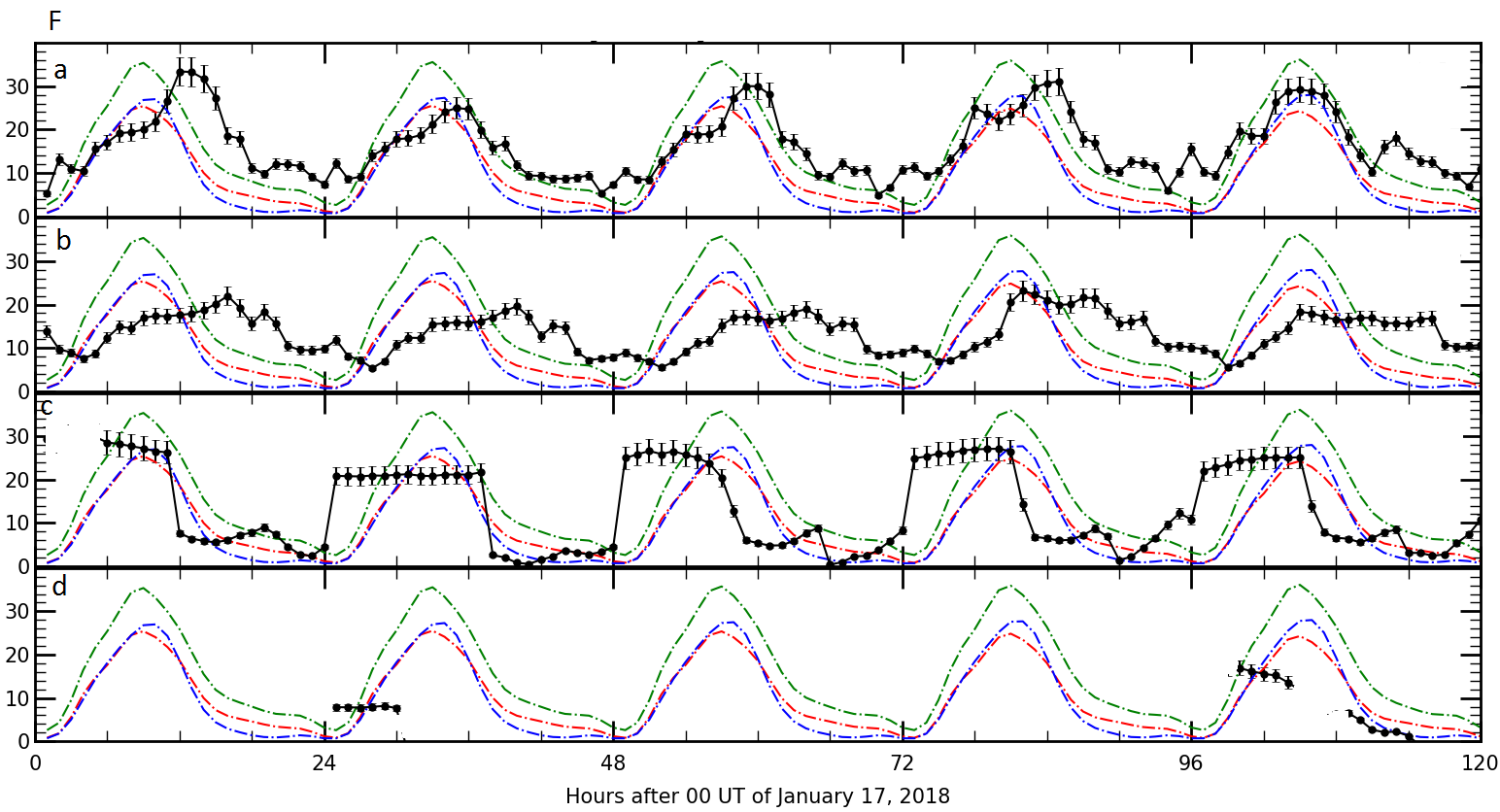}\\
\caption{Diurnal variations of model derived VTEC from IRI-P(green), NeQ(red) and IRI(blue) are compared with PRNs 2-7(panels A-F respectively) of NavIC(subplots:(a)) and all PRNs of GPS(subplots:(b)), GLONASS(subplots:(c)) and GALILEO(subplots:(d)), over Indore during January 17-21, 2018. One sigma error-bar of the measured values(black) are also shown for the period.}
\label{sc8}
\end{figure*}

Figure \ref{sc7} depicts diurnal VTEC variations obtained from IGS-GPS observables over Lucknow, Hyderabad, Bangalore and NavIC along with GPS, GLONASS and GALILEO observables over Indore. 
No significant TEC enhancements are observed on the day of Dst minimum, i.e. January 19.  
Figure \ref{sc8} shows the comparison of the three models with real time observations from NavIC, GPS, GLONASS and GALILEO VTEC over Indore in a manner similar to that explained in Figure \ref{sc4}.
It is clearly observed from Figures \ref{sc8}A-F that the models are insensitive to weak storm conditions and hence largely overestimate the real value measured during this period. The closest match is found by PRN 7 of NavIC of \ref{sc8}F.
This points out to the fact that the models are unable to capture quieter ionospheric variations around the anomaly region during a weak storm under conditions of low solar activity, when response of the ionosphere is stronger compared to higher solar activity conditions \citep{sc:40} when the solar drivers (F10.7 and SSN) of models are unable to describe the low solar EUV radiation.

\begin{figure*}
\includegraphics[width=\columnwidth,height=4.5in]{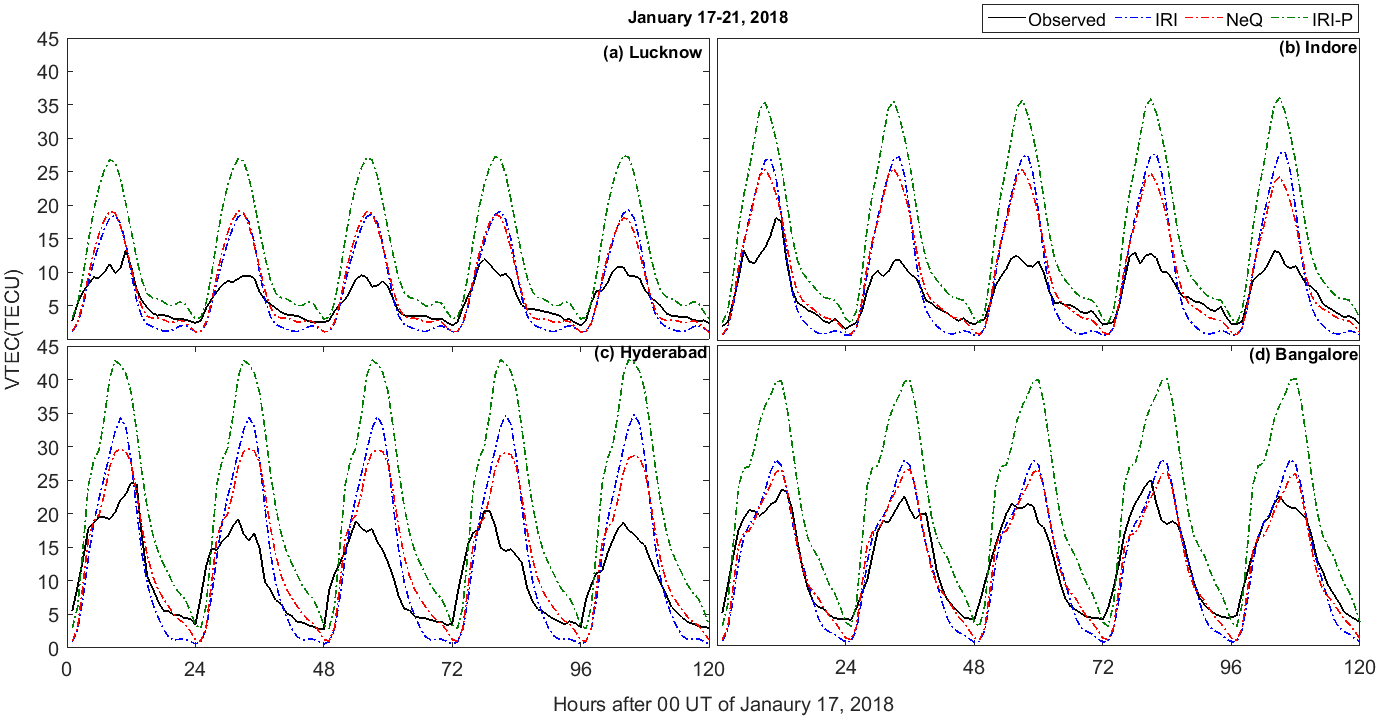}
\caption{Diurnal variations of VTEC from observed values of GPS are compared with model derived values of IRI(blue), NeQ(red) and IRI-P(green) over (a) Lucknow, (b) Indore, (c) Hyderabad and (d) Bangalore during the period of January 17-21, 2018.}
\label{sc9}
\end{figure*}

Similar to the strong storm discussed in the previous subsection, Figure \ref{sc9} depicts latitudinal variation of diurnal VTEC obtained from GPS, IRI, NeQ and IRI-P over the four stations during the weak storm period. It is clearly observed that for the stations Lucknow, Indore and Hyderabad which are near the EIA region and in between anomaly and magnetic equator respectively, the three models are largely overestimating observed values. The model TECs are also getting overestimated over Bangalore but with a lesser magnitude compared with the other stations thus further emphasizing on the inaccuracy in the models and a suggestion to improve the prediction capability, by incorporation of real measured data from GEO and GSO satellites that might produce a better picture of the ionospheric response during geomagnetic storms with moderate severity.

\newpage\subsection{Moderate storm of November 5, 2018}

\begin{figure*}
\includegraphics[width=\columnwidth,height=4in]{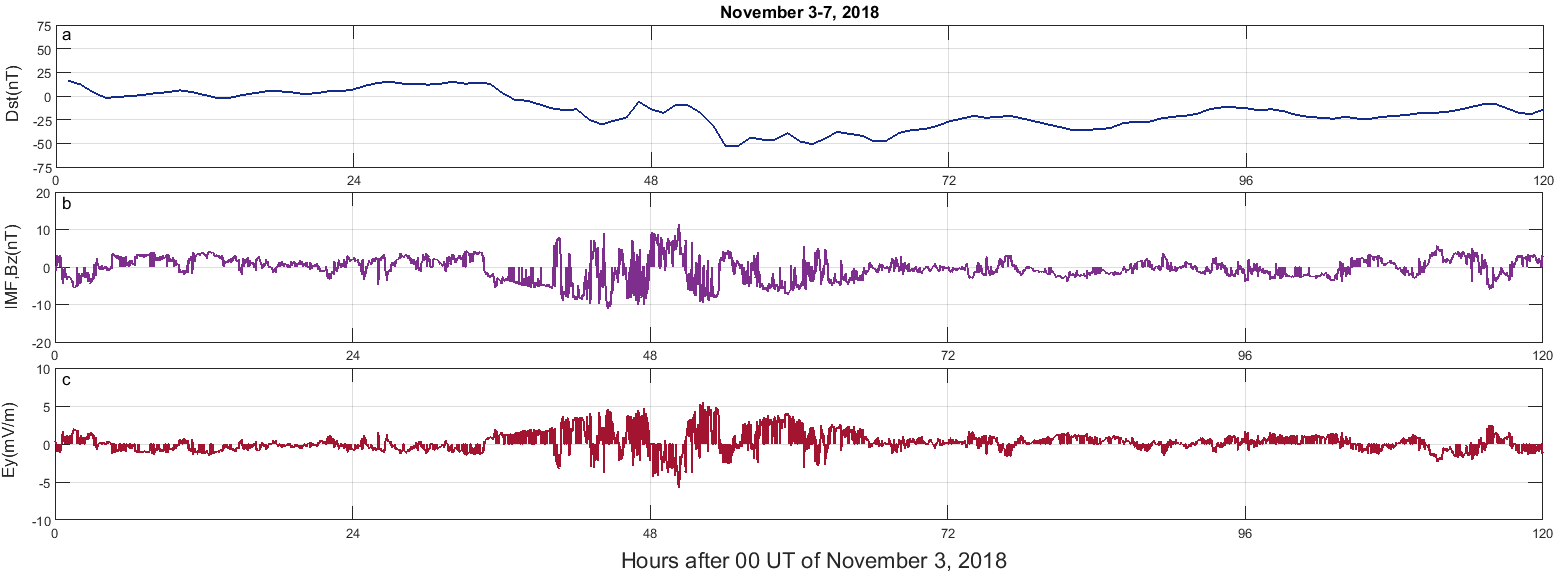}
\caption{Dst, AE and interplanetary parameters for November 3-7, 2018. November 9(48-72 UT(h) in the plot) signifies the day of Dst minimum.}
\label{sc10}
\end{figure*}

Due to a high speed stream of solar wind that made contact with the Earth's magnetic field, a G2 level (K$_p$=6, moderate) geomagnetic storm conditions were observed on November 5, 2018.
The variation in the Dst and the interplanetary parameters during this period are plotted in Figure \ref{sc10}. It is to be noted that the AE data were unavailable for this particular period. Figure \ref{sc10}a shows the development of a moderate storm as the minimum value of Dst reached to -53 nT at 06:00 UT on November 5. Figure \ref{sc10}b shows the IMF, Bz with a minimum value of -11.07 nT at 20:35 UT on November 4 while Figure \ref{sc10}c shows the IEF, Ey with a maximum value of 05.45 mV/m at 04:15 UT on November 5.

\begin{figure*}
\includegraphics[width=\columnwidth,height=4.5in]{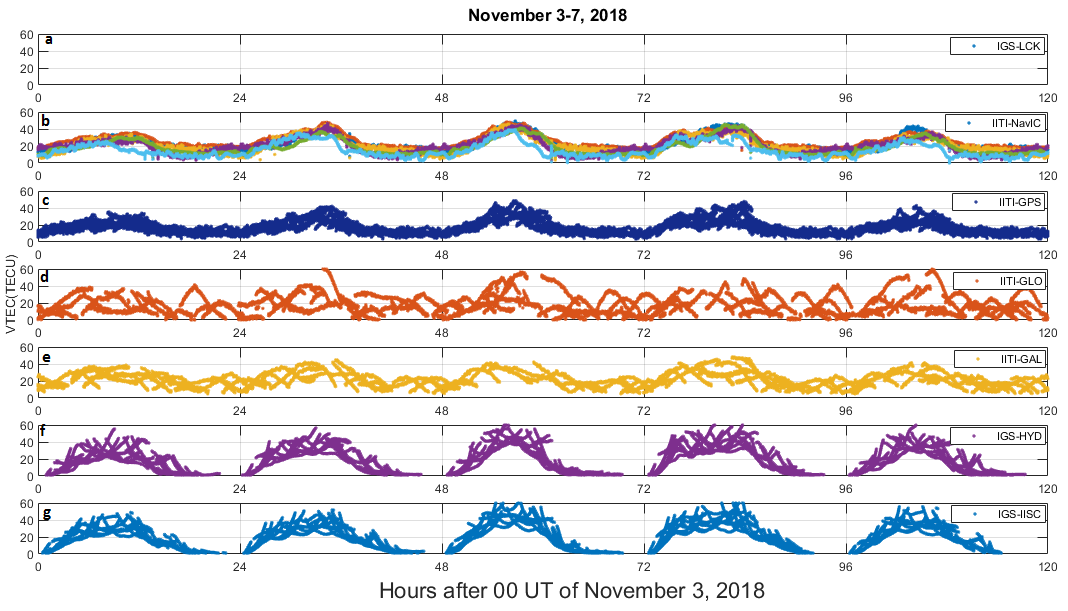}
\caption{Diurnal variation of VTEC during November 3-7, 2018 over the Indian sector.}
\label{sc11}
\end{figure*}

\begin{figure*}
\includegraphics[width=0.5\columnwidth,height=2.5in]{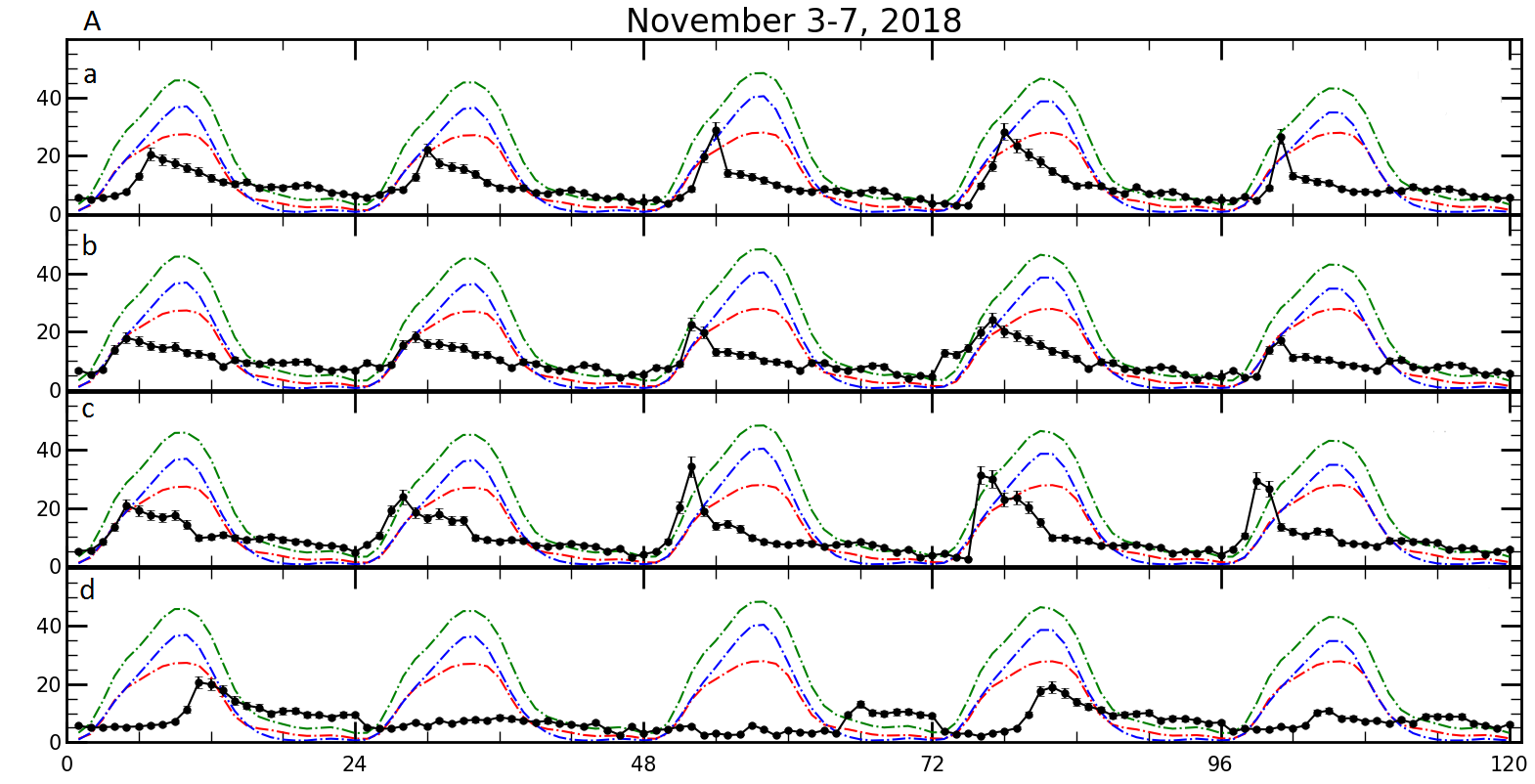}
\includegraphics[width=0.5\columnwidth,height=2.5in]{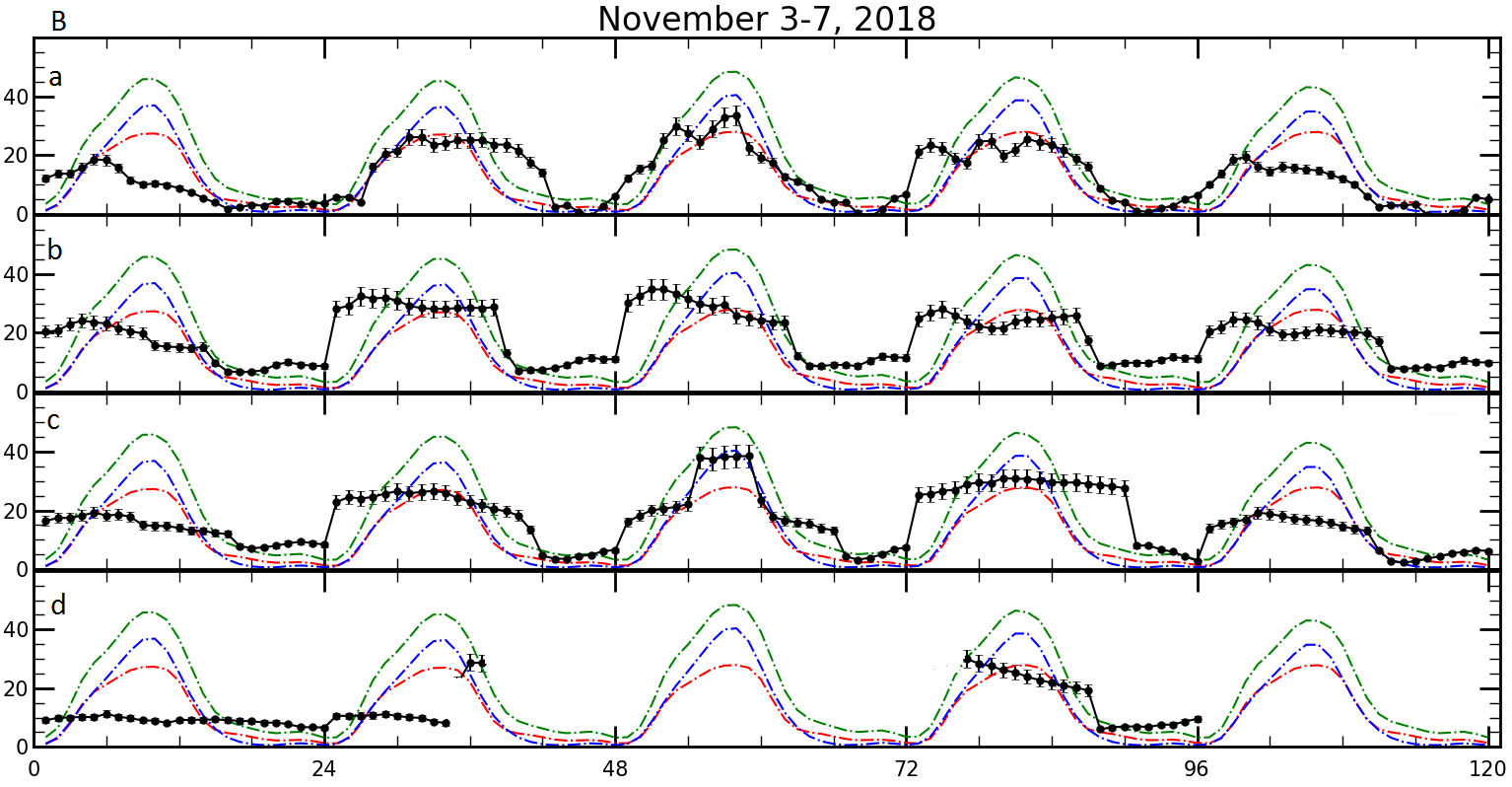}\\
\includegraphics[width=0.5\columnwidth,height=2.5in]{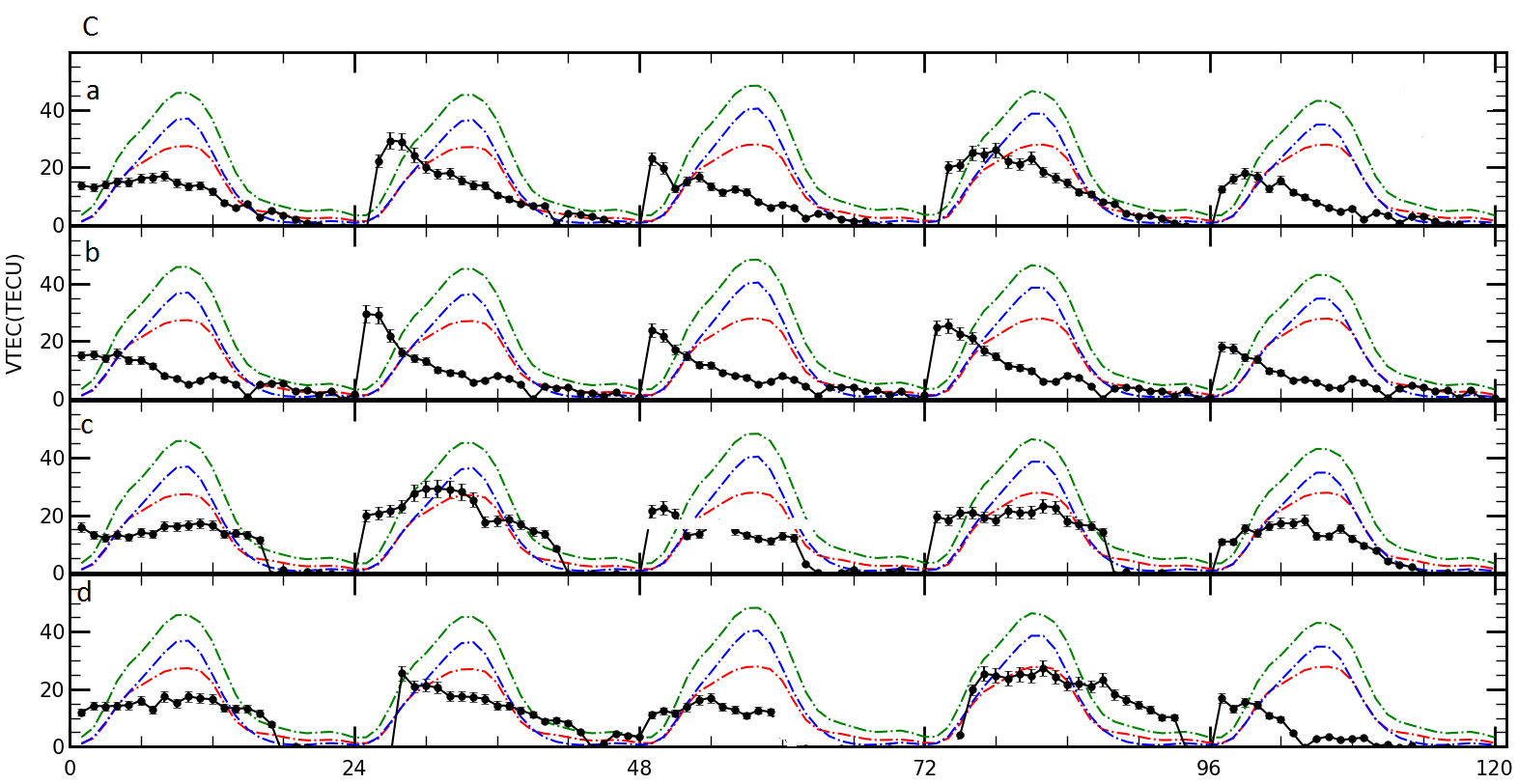}
\includegraphics[width=0.5\columnwidth,height=2.5in]{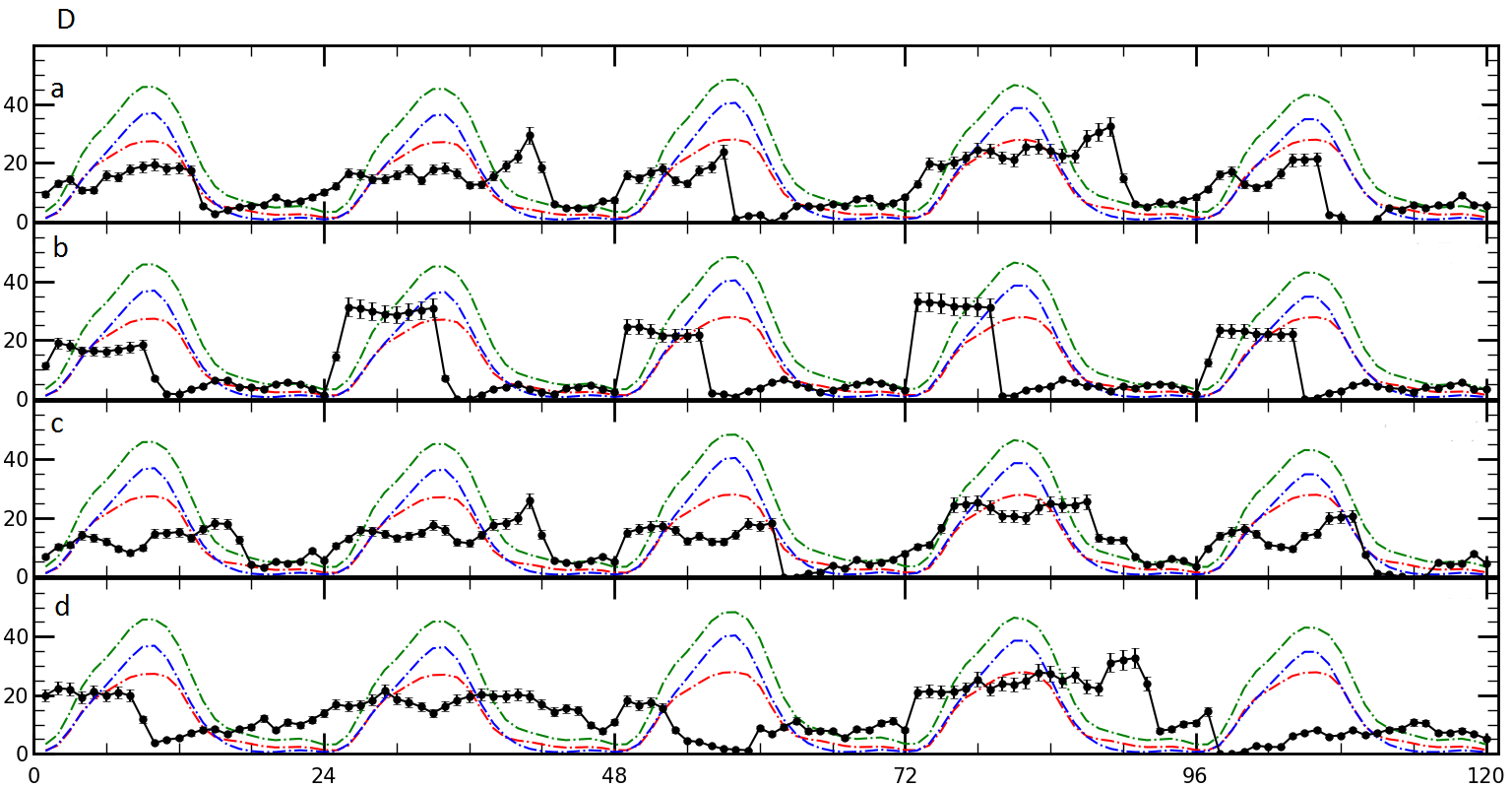}\\
\includegraphics[width=0.5\columnwidth,height=2.5in]{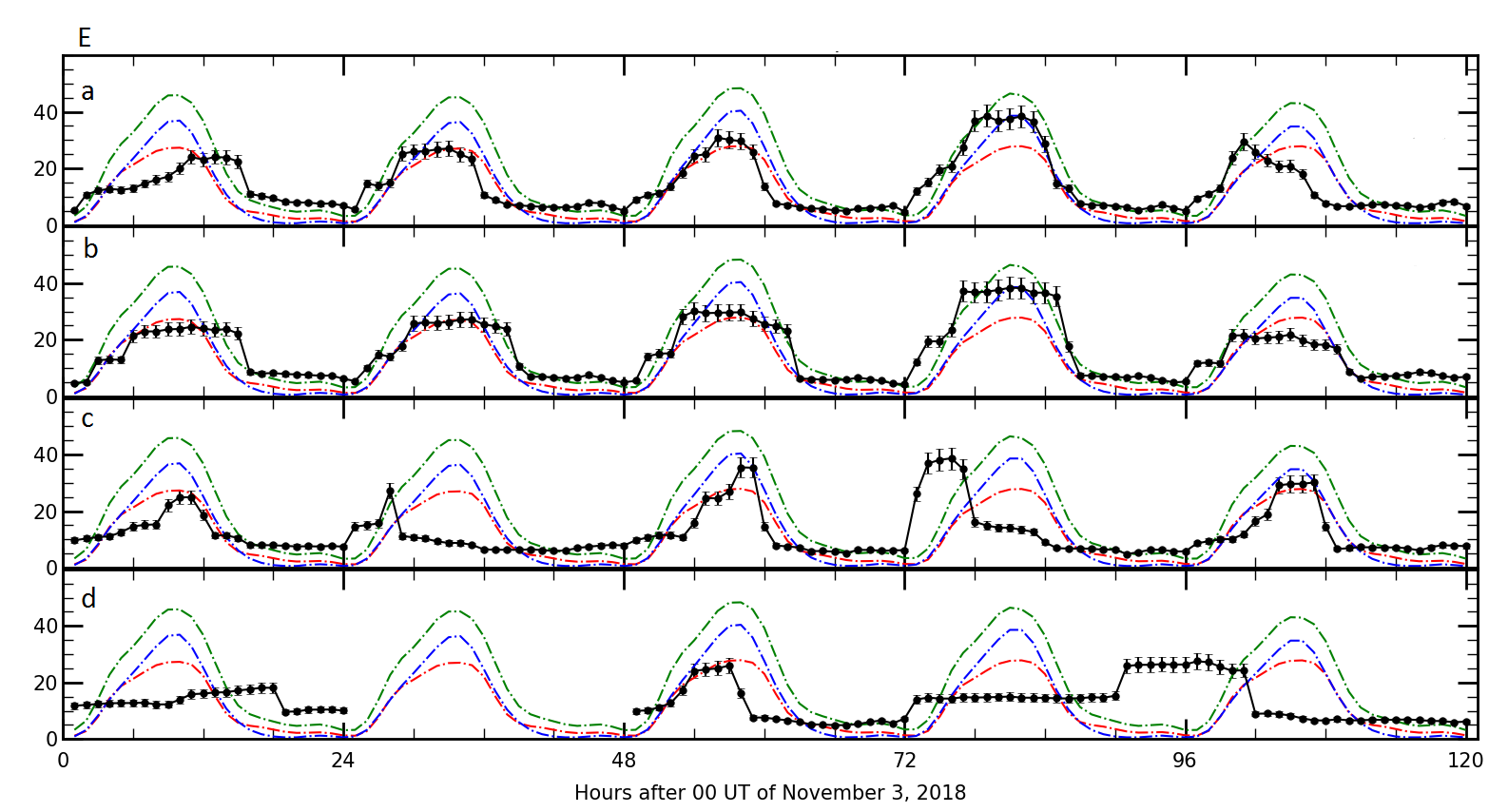}
\includegraphics[width=0.5\columnwidth,height=2.5in]{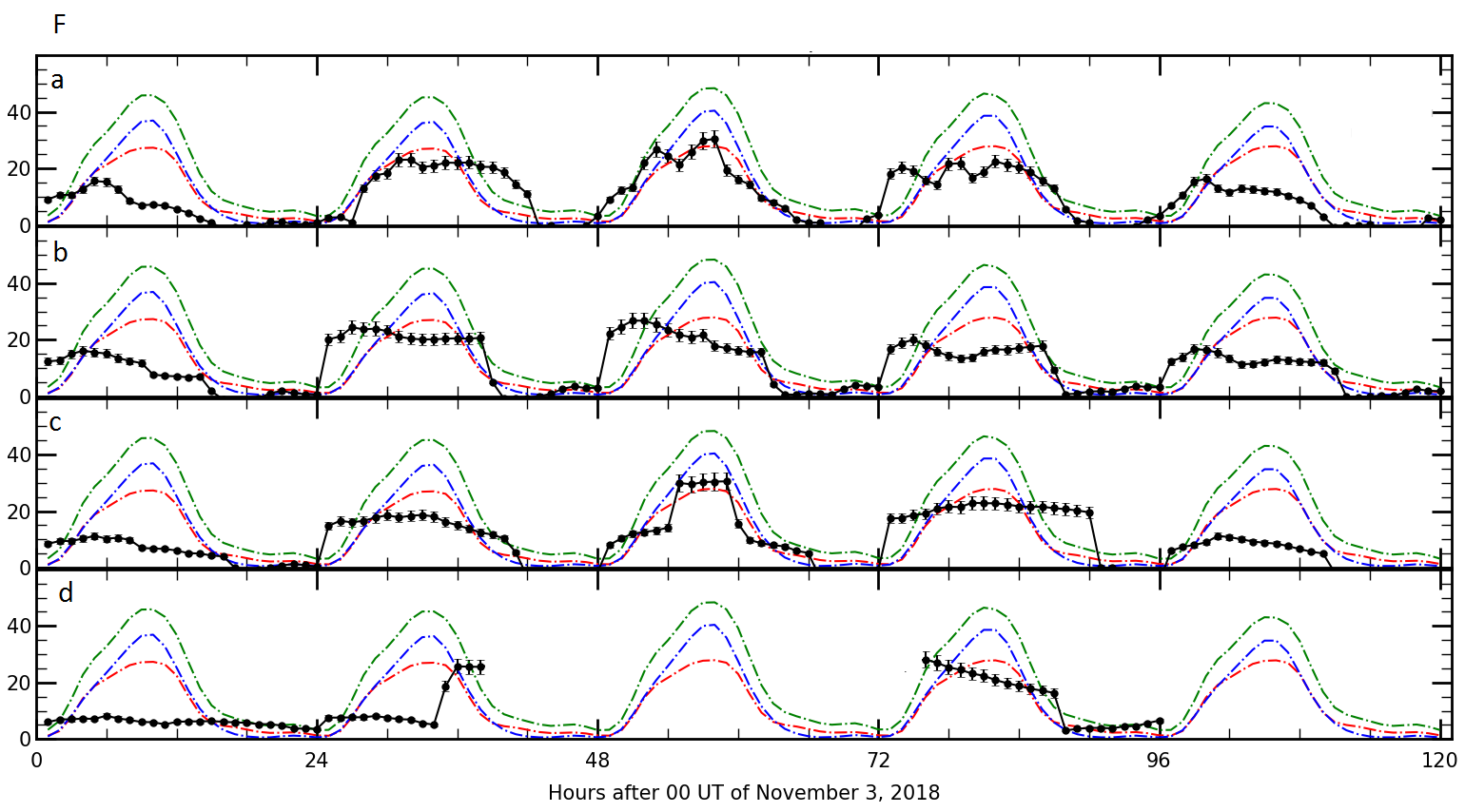}\\
\caption{Diurnal variations of model derived VTEC from IRI-P(green), NeQ(red) and IRI(blue) are compared with PRNs 2-7(panels A-F respectively) of NavIC(subplots:(a)) and all PRNs of GPS(subplots:(b)), GLONASS(subplots:(c)) and GALILEO(subplots:(d)), over Indore during November 3-7, 2018. One sigma error-bar of the measured values(black) are also shown for the period}.
\label{sc12}
\end{figure*}

\begin{figure*}
\includegraphics[width=\columnwidth,height=4.5in]{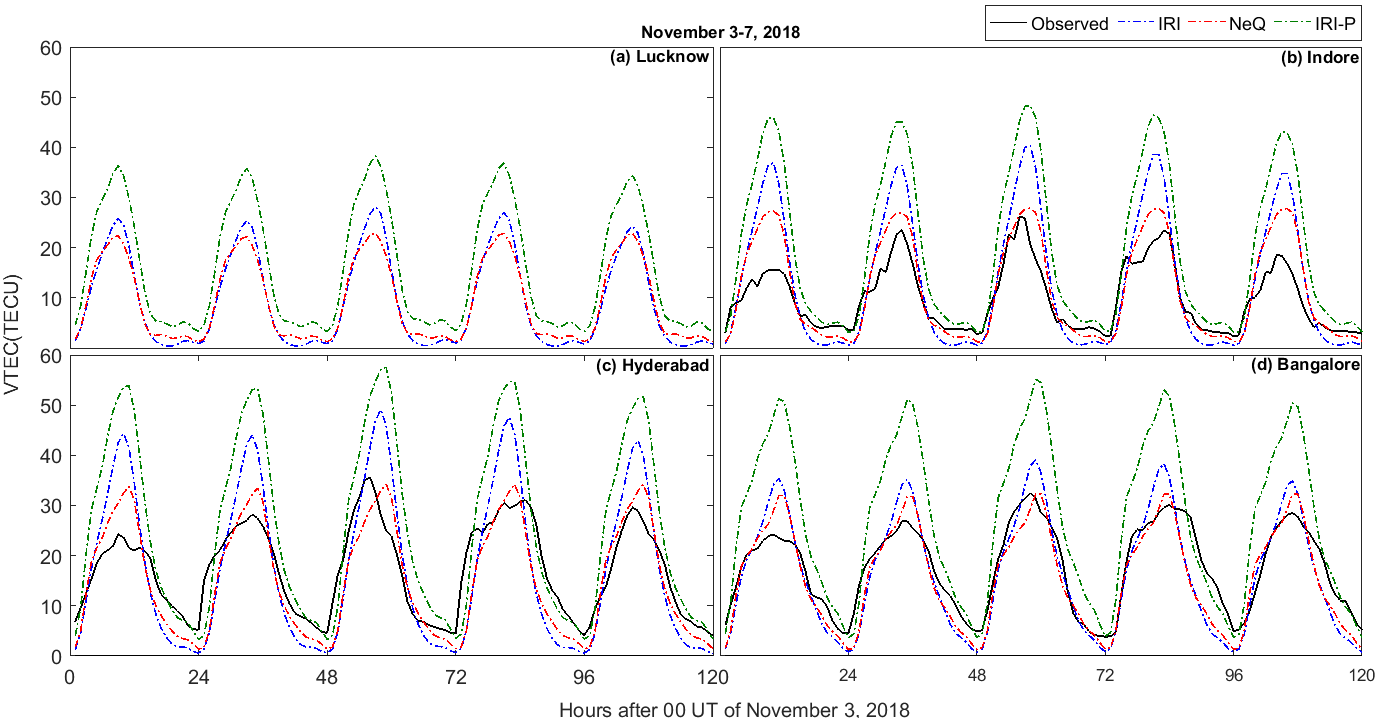}
\caption{Diurnal variations of VTEC from observed values of GPS are compared with model derived values of IRI(blue), NeQ(red) and IRI-P(green) over (a) Lucknow, (b) Indore, (c) Hyderabad and (d) Bangalore during the period of November 3-7, 2018.}
\label{sc13}
\end{figure*}

In Figure \ref{sc11}a, due to unavailability of IGS data from Lucknow during this period, it is shown as blank. No significant TEC enhancements are observed during this period from all the satellite constellations over the Indore.
Figure \ref{sc12} depicts model comparison with real time data analysed from different constellations over Indore in a manner similar to as stated in the previous two storms. 
Close match with an offset of about 5-6 TECU are observed by NavIC PRNs 3, 6 and 7(panels B, E and F respectively) and GPS values with the NeQ model. While the other PRNs' values are not at all replicated by these models. 
Thus the model predictions fail to emulate the storm time variations over this period.
Figure \ref{sc13} compares GPS TEC over Indore and IGS TEC over Hyderabad and Bangalore with the three models. Since there were no available data for Lucknow during the period, only the model values are plotted in Figure \ref{sc13}a. From the other panels, it can be observed that all the models are overestimating the measured values except over Bangalore on the storm day where almost close correspondence is shown by NeQ. Similar observations are seen from Indore on November 4-6. In all the cases significant overestimation is observed for IRI-P and IRI.  
It is expected that if the models are reliable enough around the dynamic Indian subcontinent, deviations will generally be minimum when the storms are less severe but the present study show that the models are yet to predict with precision in the low-latitude region. Therefore, due to the lack of consistency in the prediction of the variable storm time model deviations, additional data from NavIC is necessary for incorporation in the IRI-P, such that there is an improvement in the model derived predictions during geomagnetic storm time, especially over the dynamic Indian longitude sector. 

\newpage\section{Conclusions}

There has been renewed interest in the ionosphere mainly attributed to degradation in satellite navigation performance. Ionospheric variations and complexities, along with its effects on high frequency communications, has been an important field of study for decades. In order to eliminate these effects on the operation of satellite-based navigational and positioning systems (GNSS and NavIC), the existence of global ionospheric models that would provide reliable specifications of the ionospheric parameters, especially during geomagnetic storm time conditions, is essential.
In this present study, for the first time, to the best of our knowledge, the performances of the empirical models: IRI-P, NeQ, and IRI predicted TEC were compared with the NavIC and GNSS measured TEC over Indore, highlighting on the multi-constellation study over a single location near the anomaly crest. Simultaneous study of the IGS stations at Lucknow, Hyderabad, and Bangalore highlighted a single constellation study in terms of multiple locations, thereby maintaining a careful spatial distribution over the Indian longitude sector. The comparative analysis was performed under strong, moderate, and weak geomagnetic storm conditions spanning the period September 2017-November 2018 in the declining phase of solar cycle 24. 
Some correspondences, as well as inconsistencies, were observed between the measured and the model derived TEC values. 
During the strong storm of September 8, 2017, IRI-P showed the best performance in terms of matching with NavIC and GPS TEC, with an offset of about 3-5 TECU, and being able to observe the enhancement on September 7, 2017. Poor predictions were observed from all the models during the weak storm of January 2018 thereby stressing on the inaccuracy of these models during the weak storm time conditions.
The mismatch between the observed and model-derived values is attributed to the models' inherent height limitation and the difference in the topside profile estimation. The topside estimation in these models does not incorporate the effect of EIA, which leads to a poor prediction over such a dynamic region. NavIC derived values account for TEC up to geostationary, and plasmapause altitude, whereas GNSS measured values account up to $\sim$20,200 km. The TEC predicted by IRI and NeQ takes into account an altitude range upto 2000 km and GNSS altitudes, respectively, not accounting for the additional plasmaspheric electron density contribution up to NavIC altitudes ($\sim$36,000 km). 
Since IRI-P accounts for the electron density distribution up to plasmaspheric heights, TEC values predicted by it are expected to be higher and closer to the NavIC satellite measured values. The NavIC satellites located at higher altitudes would provide a better measure of the plasmaspheric contributions in addition to the continuous monitoring of TEC available from a single satellite of NavIC. Thus a possible suggestion could be the inclusion of NavIC measured values to the database of IRI-P in order to achieve model predictions with greater accuracy in locations around the Indian longitude sector.

\newpage
\section*{Acknowledgments} 

SC acknowledges Space Applications Centre, ISRO for providing research fellowship under the project number: NGP-17 of the GAGAN/NavIC Utilization Program. DA acknowledges Department of Science and Technology, India for providing INSPIRE Fellowship. The authors also acknowledge Prof. Gopi Krishna Seemala of the Indian Institute of Geomagnetism, Navi Mumbai, India for providing the software to analyze the IGS data (available at http://sopac.ucsd.edu/dataBrowser.shtml). SC would like to thank Nivedita Chakraborty and Aishrila Mazumder for fruitful discussions. Acknowledgements go to World Data Center for Geomagnetism, Kyoto for the hourly Dst and $K_p$ index (available at http://wdc.kugi.kyoto.ac.jp), the Sunspot Index and Long-term Solar Observations (SILSO) in Belgium (available at http://www.sidc.be
/silso/datafiles) and NASA's SPDF omniweb service (available at https:
//omniweb.gsfc.nasa.gov
/form/omni\_min.html) for the solar radio flux, the AE index and the interplanetary magnetic and electric field data. 
Acknowledgements are also given for the model development of the IRI, NeQ and IRI-P models freely available at https://ccmc.gsfc.nasa.gov/
modelweb/models/iri2016\_vitmo.php, https://t-ict4d.ictp.it/nequick2/nequick-2-web-model and http://www.ionolab.org respectively. Finally, the authors acknowledge the anonymous reviewers for helpful suggestions that improved quality of the manuscript.

\newpage
\bibliographystyle{model2-names.bst}\biboptions{authoryear}

\bibliography{article.bib}

\end{document}